\documentclass[prd,nofootinbib,preprint,superscriptaddress,floatfix]{revtex4-1}

\usepackage{amsmath, amssymb, graphicx}

\usepackage{xcolor}
\definecolor{revisionblue}{RGB}{0,0,255}
\definecolor{REVISIONBLUE}{RGB}{0,0,255}
\definecolor{blue}{RGB}{0,0,255}
\definecolor{BLUE}{RGB}{0,0,255}
\definecolor{revisionorange}{RGB}{230,95,0}

\newcommand{\eq}{\mathrm{eq}}
\newcommand{\be}{\begin{equation}}
\newcommand{\ee}{\end{equation}}
\newcommand{\bea}{\begin{eqnarray}}
\newcommand{\eea}{\end{eqnarray}}

\newcommand{\etaB}{\eta_B}

\newcommand{\dd}{\mathop{}\!\mathrm{d}}
\newcommand{\SM}{\mathrm{SM}}
\newcommand{\PBH}{\mathrm{PBH}}
\newcommand{\BH}{\mathrm{BH}}
\newcommand{\Therm}{\mathrm{th}}

\newcommand{\gstarS}{g_{\star s}}

\begin{document}
\title{Primordial Black Hole Assisted Dirac Leptogenesis}

\author{Marco Manno}
\email{marco.manno@unisalento.it}

\affiliation{%
Dipartimento di Matematica e Fisica ``Ennio De Giorgi,'' Universit\`a del Salento, 73100 Lecce, Italy
}
\affiliation{%
INFN-Istituto Nazionale di Fisica Nucleare, Sezione di Lecce, 73100 Lecce, Italy
}

\author{Anish Ghoshal}
\email{A.Ghoshal@sussex.ac.uk}
\affiliation{Department of Physics and Astronomy, University of Sussex, \\
Brighton, BN1 9RH, United Kingdom}

\begin{abstract}
Dirac leptogenesis offers a qualitatively different route to the baryon asymmetry: total lepton number is conserved, but equal and opposite asymmetries are stored in the Standard Model and right handed neutrino sectors, with only the visible component processed by electroweak sphalerons.  We present the first dedicated study of how evaporating primordial black holes (PBHs) modify this mechanism.  In a minimal charged-scalar mediator setup, we solve the coupled Boltzmann system including thermal production, Hawking emission of the mediator $X$ and of right handed neutrinos, the PBH contribution to the expansion rate, and entropy injection.  We find that Hawking emission can populate $X$ in regions where thermal production is inefficient, allowing its subsequent CP-violating decays to enhance the baryon asymmetry. For representative benchmarks, PBHs extend the successful region from mediator masses around $10^9\,{\rm GeV}$ down to about $10^6\,{\rm GeV}$ and can compensate for smaller effective CP asymmetries. This effect is not universal: where thermal Dirac leptogenesis is already efficient, PBHs can instead reduce the final asymmetry. Their evaporation also deposits energy in the right handed neutrino sector and contributes to $\Delta N_{\rm eff}$.  In the PBH parameter scan, part of the region that reproduces the observed baryon asymmetry lies below the Planck reference level $\Delta N_{\rm eff}=0.284$ and above the projected CMB-S4 sensitivity $\Delta N_{\rm eff}=0.06$.
\end{abstract}

\maketitle

\tableofcontents

\section{Introduction}

The observed matter-antimatter asymmetry is one of the most direct pieces of evidence that the early Universe contained physics beyond thermal equilibrium in the Standard Model.  Any dynamical explanation must satisfy the Sakharov conditions: baryon number violation, $C$ and $CP$ violation, and departure from thermal equilibrium~\cite{Sakharov:1967dj}.  Leptogenesis is especially compelling because it connects this cosmological asymmetry to neutrino physics: a lepton asymmetry generated in the early Universe is partially converted into baryon number by electroweak sphalerons before they freeze out~\cite{Fukugita:1986hr,PhysRevD.30.2212,Kuzmin:1985mm,PhysRevD.36.581,Harvey:1990qw,DOnofrio:2014rug}.  In this sense, baryogenesis is not merely a cosmological question; it is also a probe of the origin of neutrino mass.

Most realizations of leptogenesis assume that neutrinos are Majorana fermions, as in the original high-scale scenario~\cite{Fukugita:1986hr}.  This is elegant and predictive, but it is not experimentally established.  No lepton-number-violating signal has yet been observed, and the possibility that neutrinos are Dirac particles remains fully viable.  Dirac leptogenesis is therefore not a minor variant of the standard picture, and it has also been explored in radiative neutrino-mass and dark-sector setups, including scotogenic Dirac constructions~\cite{Borah:2016zbd}.  It is the logically distinct possibility that the baryon asymmetry was generated while total lepton number, or equivalently $B-L$, remained conserved.  The mechanism relies on equal and opposite asymmetries stored in the Standard Model and right handed neutrino ($\nu_R$) sectors: sphalerons process only the left-handed component, while the tiny Dirac Yukawa couplings keep the right handed sector out of equilibrium until after sphaleron freeze-out~\cite{Dick:1999je,Murayama:2002je,Gu:2006dc,Gu:2007mc,Narendra:2017uxl,Heeck:2013vha,Gu:2019yvw,Mahanta:2021plx,Heeck:2023soj}.  This makes Dirac leptogenesis an important target in its own right, because it links the baryon asymmetry to the non-thermalization of otherwise elusive right handed neutrinos and offers a cosmological way to test the Dirac-neutrino hypothesis.
The range of known Dirac-leptogenesis mechanisms has expanded substantially.  In the wash-in construction of Ref.~\cite{Blazek:2024efd}, reactions among the light plasma degrees of freedom build the $\nu_R$ asymmetry without requiring an on-shell thermal population of the heavy mediator.  Other recent realizations place the asymmetry source in vectorlike-lepton decays within a left-right-symmetric theory~\cite{Babu:2024lrsm}, or in the reheating era, with an initially asymmetric scalar sector subsequently communicating its charge to the two neutrino chiralities~\cite{Ahmed:2025primordial}.  These constructions show that Dirac leptogenesis can be realized through several distinct mechanisms.  A further variant links the same conserved-charge bookkeeping to asymmetric dark matter by distributing the primordial charge between visible and dark sectors~\cite{Ishida:2025adm}.  These developments differ in field content and asymmetry-transfer dynamics from the mechanism studied here.  The present work addresses the distinct question of how Hawking production modifies the charged-scalar decay realization of Ref.~\cite{Heeck:2023soj}.

The same feature that makes Dirac leptogenesis attractive also makes it demanding.  The asymmetry must be produced efficiently enough before sphalerons freeze out, yet the right handed neutrinos must remain sufficiently decoupled.  In minimal charged-mediator realizations this requirement depends on the interplay among heavy scalar decays, inverse decays, washout, annihilations and the timing of the electroweak transition~\cite{Heeck:2023soj,Blazek:2024efd}.  In the standard thermal history, these competing requirements can restrict the viable parameter space and push the relevant mediator scale high.  A modified early Universe can therefore alter the viable parameter space qualitatively.

Primordial black holes (PBHs) provide a particularly sharp and well-motivated example of such a modified cosmological history.  PBHs can form from large primordial density fluctuations or related early-Universe mechanisms~\cite{Carr:1974nx,Carr:1975qj,Sasaki:2018dmp,Carr:2020gox}.  If sufficiently light, they evaporate by Hawking radiation long before the present epoch, injecting entropy, modifying the Hubble rate, and producing particles independently of their Standard Model gauge charges~\cite{Hawking:1974rv,Hawking:1975vcx,Page:1976df,Page:1977um,MacGibbon:1991tj,Cheek:2021odj}.  This gravitational production channel can populate heavy states even when the thermal bath is too cold to produce them efficiently.  PBH evaporation has therefore been studied as a source of baryogenesis and as a tool for modifying leptogenesis in several settings~\cite{Turner:1979bt,Baumann:2007yr,Hooper:2020otu,Fujita:2014hha,Morrison:2018xla,Perez-Gonzalez:2020vnz,JyotiDas:2021shi,Bernal:2022pue,Barman:2022gjo,Calabrese:2023bxz,Schmitz:2023pfy,Calabrese:2023key,Ghoshal:2023fno,Barman:2024slw,Gunn:2024xaq,Calabrese:2025sfh,Borah:2024memburden,Borah:2026zbl}.
Among the nearest PBH analogues, Ref.~\cite{Barman:2023colored} uses Hawking-emitted colored scalars followed by baryon-number-violating decays, whereas Ref.~\cite{Schmitz:2023pfy} uses PBHs to repopulate heavy asymmetry-generating states and then converts the resulting SM charge background through wash-in.  A conceptually different possibility biases the Hawking flux itself through an effective chemical potential at the horizon, so that no secondary unstable parent is needed to create baryon number~\cite{Iguaz:2025asymmetric}.  All of these scenarios exploit late gravitational production, but the charge flow in our setup is different.  The black holes emit a charged scalar symmetrically; lepton number remains exact, and only its subsequent CP-asymmetric branching divides the compensating charges between the SM bath and $\nu_R$.

We consider the minimal charged-scalar mediator setup of Ref.~\cite{Heeck:2023soj} and solve the coupled Boltzmann system including thermal production, Hawking production of the charged scalar $X$, direct Hawking emission of $\nu_R$, the PBH contribution to the expansion rate, and entropy injection.  The central physical point is simple: PBHs can create a nonthermal population of the mediator $X$ at times when the thermal population is inefficient or already depleted.  The subsequent CP-violating decays of this PBH-produced mediator provide an additional source for the stored Dirac asymmetry.

The basic analytical expectation follows from two observations.  First, heavy-particle emission becomes efficient when the Hawking temperature reaches the relevant mass scale, $T_{\rm BH}\sim M_X$.  Second, during radiation domination, the PBH energy fraction grows approximately with the scale factor until evaporation, so even a small initial abundance can substantially modify the subsequent evolution.  These observations explain the structure of the analysis.  The $(M_{\rm PBH},\beta')$ plane determines when and how strongly the nonthermal source appears, while the particle-physics plane determines whether the emitted mediator can convert this source into a surviving baryon asymmetry.

Our main result is that PBHs can rescue regions where thermal Dirac leptogenesis is inefficient.  Within the numerical framework and approximations used below, representative benchmarks show that Hawking production lowers the required charged-mediator scale from roughly $10^9\,{\rm GeV}$ to $10^6\,{\rm GeV}$ and allows successful baryogenesis for smaller effective CP asymmetries.  The effect is not a universal enhancement: in regions where the thermal asymmetry is already efficient, PBHs can instead dilute or sign-reshape the final result.  Moreover, the same PBH evaporation history populates the right handed neutrino sector and therefore produces a correlated dark-radiation signal, quantified by $\Delta N_{\rm eff}$.  PBH-assisted Dirac leptogenesis can therefore extend the viable baryogenesis parameter space while producing a correlated target for precision CMB measurements.

The paper is organized as follows.  In Sec.~\ref{sec:dirac_review}, we review the charged-mediator realization of Dirac leptogenesis and the corresponding thermal Boltzmann equations.  In Sec.~\ref{sec:pbh_review}, we summarize the PBH formation and evaporation ingredients used in the calculation.  In Sec.~\ref{sec:pbh_dirac}, we present the coupled PBH-modified Boltzmann system, including the energy-density evolution and the dark-radiation observable.  In Sec.~\ref{sec:results}, we present the numerical evolution and the scans over the PBH and particle parameters.  Section~\ref{sec:majorana_comparison} compares our results with PBH-assisted Majorana leptogenesis, and Sec.~\ref{sec:gw_probes} discusses possible gravitational-wave probes.  We conclude in Sec.~\ref{sec:conclusion}.  Appendix~\ref{app:background_evolution} gives the background energy budget shown in Fig.~\ref{fig:energy_fractions_appendix}; Appendix~\ref{app:signed_eta} displays the sign of $\eta_B$ and clarifies the sign-changing structures in the particle-parameter scans through Fig.~\ref{fig:signed_eta_appendix}; Appendix~\ref{app:branching_ratio} tests the complementary branching-ratio choice in Fig.~\ref{fig:branching_ratio_appendix}; and Appendix~\ref{app:gw_details} collects the gravitational-wave estimates.

\section{Dirac Leptogenesis: Review} \label{sec:dirac_review}
The defining feature of Dirac leptogenesis is that the cosmological evolution preserves total lepton number, while
equal and opposite lepton asymmetries are generated in the
Standard Model (SM) and right handed neutrino ($\nu_R$) sectors~\cite{Dick:1999je,Murayama:2002je}.
The small Dirac Yukawa couplings of the light neutrinos keep $\nu_R$
out of equilibrium until after electroweak sphaleron freeze-out.  Sphalerons
therefore act only on the left-handed SM plasma and convert part of its charge
into baryon number.

We consider case $a$ of Ref.~\cite{Heeck:2023soj}, following the notation and classification of the minimal charged-mediator realizations discussed there.\footnote{We focus on case
$a$ because it provides the simplest realization. In the thermal analysis of
Ref.~\cite{Heeck:2023soj}, most of the other mediator assignments lead to
qualitatively similar phenomenology, apart from the leptoquark cases $c$ and
$d$.}  In addition to the three
right handed neutrinos $\nu_{R\alpha}$, the SM is supplemented by at least two
electrically charged scalar mass eigenstates
\begin{equation}
  X_i\equiv X_i^-\sim(\mathbf{1},\mathbf{1},-1),
  \qquad \bar X_i\equiv X_i^+,
  \label{eq:X_quantum_numbers_review}
\end{equation}
where the entries give the $SU(3)_c\times SU(2)_L\times U(1)_Y$ quantum
numbers in the convention $Q=T_3+Y$.  Since $X_i$ is an $SU(2)_L$ singlet,
its hypercharge equals its electric charge.  $X_i$ and $\bar X_i$
denote the two individual charge states of the same complex scalar, with
$g_X=1$ for each charge state and two scalar degrees of freedom for the
$X_i+\bar X_i$ system.  The relevant interactions are
\begin{equation}
  \mathcal L_X=
  \frac12 (F_i)_{\alpha\beta}\,
  \overline{L_\alpha^c}i\sigma_2L_\beta\,\bar X_i
  +(G_i)_{\alpha\beta}\,
  \overline{e^c_{R\alpha}}\nu_{R\beta}\,\bar X_i
  +{\rm h.c.}
  \label{eq:dirac_lagrangian_review}
\end{equation}
Here $L_\alpha$ and $e_{R\alpha}$ are the SM lepton doublets and right handed
charged lepton singlets, $\alpha,\beta$ are flavor indices, $\sigma_2$ is the
second Pauli matrix, and the superscript $c$ denotes charge conjugation.  The
antisymmetric $SU(2)_L$ contraction implies $F_i^T=-F_i$, whereas $G_i$ is a
general complex matrix.  The interaction terms displayed in
Eq.~\eqref{eq:dirac_lagrangian_review} contain $\bar X_i=X_i^+$; their Hermitian
conjugates contain $X_i=X_i^-$ and describe the decay channels written below.

Assigning $L(X_i)=2$ makes both interactions conserve total lepton number.  A
conserved $B-L$ symmetry forbids Majorana masses.  The simultaneous presence
of $F_i$ and $G_i$ breaks
the separate $\nu_R$ number and permits an asymmetry to be generated in the
$X_i$ decays.  At least two scalar mass eigenstates are required for a nonzero CP
asymmetry through loop interference.  We denote the two inclusive decay
classes by $X_i\to e_R\nu_R$ and $X_i\to e_L\nu_L$, with the corresponding
charge-conjugate decays of $\bar X_i$ understood.

Neglecting final-state masses and summing over flavors, the tree-level partial widths for one charge state are
\begin{equation}
  \Gamma(X_i\to e_R\nu_R)=\frac{M_i}{16\pi}{\rm Tr}(G_iG_i^\dagger),
  \qquad
  \Gamma(X_i\to e_L\nu_L)=\frac{M_i}{16\pi}{\rm Tr}(F_iF_i^\dagger).
  \label{eq:tree_rates_review}
\end{equation}
Here $M_i$ is the mass of $X_i$.  Writing
$\Gamma_i\equiv\Gamma(X_i)=\Gamma(\bar X_i)$ for the total width of one charge
state, we define
\begin{align}
  B_R^i&=\frac{\Gamma(X_i\to e_R\nu_R)+\Gamma(\bar X_i\to \bar e_R\bar\nu_R)}{2\Gamma_i},
  &
  B_L^i&=\frac{\Gamma(X_i\to e_L\nu_L)+\Gamma(\bar X_i\to \bar e_L\bar\nu_L)}{2\Gamma_i},
  \\
  \epsilon_i&=\frac{\Gamma(X_i\to e_R\nu_R)-\Gamma(\bar X_i\to \bar e_R\bar\nu_R)}{2\Gamma_i}.
  \label{eq:branching_epsilon_review}
\end{align}
Here $B_L^i+B_R^i=1$.  Since these are the only two decay classes, CPT equality of the total widths implies an equal and opposite CP asymmetry in the left-handed channel.  At this stage, the index $i$ labels a generic mediator, and the necessary positivity bound $|\epsilon_i|\leq\min(B_L^i,B_R^i)$ applies separately to each state~\cite{Heeck:2023soj}.  We do not impose the additional constraints associated with a specific flavor realization.
The positivity inequality is necessary because each partial width must remain non-negative, but it is not sufficient to guarantee that a chosen $(B_L^i,B_R^i,\epsilon_i)$ can be generated by a common set of perturbative flavor matrices and scalar masses\footnote{  Accordingly, Fig.~\ref{fig:particle_plane_etaB} should be read as an effective-parameter reach of the Boltzmann system, not as a scan in which every point has been demonstrated to arise from a complete ultraviolet flavor model.}.

Following Ref.~\cite{Heeck:2023soj}, we assume a hierarchical scalar spectrum
with $X_1$ as the lightest state.
The CP asymmetry $\epsilon_1$ arises from the interference between the
tree-level decay of $X_1$ and one-loop diagrams involving the heavier scalars
$X_{j>1}$.  For a hierarchical scalar spectrum~\cite{Heeck:2023soj},
\begin{equation}
  \epsilon_1\simeq
  \frac{\displaystyle\sum_{j>1}
  \left(M_j^2/M_1^2-1\right)^{-1}
  \operatorname{Im}\!\left[
  \operatorname{Tr}(F_1F_j^\dagger)
  \operatorname{Tr}(G_1G_j^\dagger)\right]}
  {8\pi\left[
  \operatorname{Tr}(G_1G_1^\dagger)+
  \operatorname{Tr}(F_1F_1^\dagger)\right]}\,.
  \label{eq:epsilon_parametric_revision}
\end{equation}
This expression shows explicitly that a nonzero CP asymmetry requires both classes of decay
couplings, at least one heavier scalar and nontrivial CP phases.
Equation~\eqref{eq:epsilon_parametric_revision} is the hierarchical, zero-width limit of the one-loop interference result quoted in Ref.~\cite{Heeck:2023soj}.  In this minimal charged-scalar realization the relevant CP-odd invariant is generated by mixing, or wave-function, interference between $X_1$ and the heavier charged scalars.  The expression does not include a resonant resummation; it is therefore intended for mass splittings large compared with the scalar widths.  Possible finite-width effects and additional flavor-dependent contributions are not modeled explicitly in Eq.~\eqref{eq:epsilon_parametric_revision} or correlated with the parameters of the subsequent effective scan.
  As in
Ref.~\cite{Heeck:2023soj}, we parameterize CP violation through $\epsilon$
without selecting a particular flavor realization and impose the general
bound given above.

To relate the generated lepton asymmetry to the baryon abundance, we use
entropy-normalized yields and define
\begin{equation}
  Y_A\equiv \frac{n_A}{s},
  \qquad
  Y_{\Delta_{\nu_R}}\equiv Y_{\nu_R}-Y_{\bar\nu_R}.
\end{equation}
Here $n_A$ is the physical number density of species $A$, $s$ is the entropy
density of the SM thermal bath, and $Y_{\nu_R}$ sums over the three
right handed neutrino flavors $\nu_{R\alpha}$, with $\alpha=1,2,3$.  With this
sign convention, the baryon yield is
\begin{equation}
  Y_B=\chi_{\rm sph}Y_{\Delta_{\nu_R}}(T_{\rm sph}),
  \qquad \chi_{\rm sph}=\frac{28}{79},
  \qquad T_{\rm sph}=131.7\,{\rm GeV},
  \label{eq:sphaleron_conversion_review}
\end{equation}
where $\chi_{\rm sph}$ is the sphaleron conversion
factor~\cite{Harvey:1990qw}, while the quoted temperature is the central value
of the SM sphaleron freeze-out result~\cite{DOnofrio:2014rug}. The PBHs considered in this work evaporate well before the electroweak epoch, so we use the standard value of $T_{\rm sph}$.
The positive sign in Eq.~\eqref{eq:sphaleron_conversion_review} follows from our definition $\Delta_{\nu_R}=Y_{\nu_R}-Y_{\bar\nu_R}$.  Starting from vanishing total $B-L$, conservation of total $B-L$ implies $(B-L)_{\rm SM}=\Delta_{\nu_R}$ while the right handed neutrinos remain chemically decoupled.  The standard SM spectator relation $B=(28/79)(B-L)_{\rm SM}$ then gives Eq.~\eqref{eq:sphaleron_conversion_review}.  Definitions in which the visible lepton asymmetry, rather than the right handed neutrino asymmetry, is evolved commonly display an additional minus sign.
  In the Boltzmann equations we retain only $X_1$.\footnote{We apply the same
lightest-state approximation to the Hawking emission discussed in
Sec.~\ref{sec:pbh_review}.  Production of the heavier scalars is assumed to be
suppressed by their larger masses; including it would require extending the
Boltzmann system to the additional mediator populations.}  From this point
onward we write
$X\equiv X_1$, $M_X\equiv M_1$, $\Gamma_X\equiv\Gamma_1$,
$B_{L,R}\equiv B_{L,R}^1$ and $\epsilon\equiv\epsilon_1$.  Although their
abundances are not evolved, the heavier states enter through the loop
contribution to $\epsilon$ in Eq.~\eqref{eq:epsilon_parametric_revision}.  As
noted in Ref.~\cite{Heeck:2023soj}, any asymmetry generated by these states is
expected to be washed out by the interactions of $X$, while their decays
would increase the $\nu_R$ abundance and hence $\Delta N_{\rm eff}$.
We define
\begin{equation}
  \Sigma_A\equiv Y_A+Y_{\bar A},
  \qquad
  \Delta_A\equiv Y_A-Y_{\bar A},
  \label{eq:sigma_delta_review}
\end{equation}
and use the same convention for the corresponding equilibrium abundance
$\Sigma_A^{\rm eq}$, which includes both particles and antiparticles.
In particular, $\Sigma_X$ and $\Sigma_X^{\rm eq}$ include both
charge states.
For the right handed neutrinos, $Y_{\nu_R}$ and $Y_{\bar\nu_R}$ denote the
abundances summed over the three flavors.  Thus, the three $\nu_R$ flavors
enter only through $\Sigma_{\nu_R}$ and $\Delta_{\nu_R}$, which describe the
total abundance and asymmetry of the right handed neutrino sector.  Production
and redistribution among individual flavors are not followed separately.
The corresponding Boltzmann equations are~\cite{Heeck:2023soj}
\begin{align}
\frac{d\Sigma_X}{dy}&=\frac12\langle\sigma v\rangle_{X\bar X}(\Sigma_X^2-\Sigma_X^{{\rm eq}2})+\frac{\langle\Gamma_X\rangle}{s}\left[\Sigma_X-\Sigma_X^{\rm eq}\left(B_L+B_R\frac{\Sigma_{\nu_R}}{\Sigma_{\nu_R}^{\rm eq}}\right)\right],\\
\frac{d\Sigma_{\nu_R}}{dy}&=-\frac{\langle\Gamma_X\rangle}{s}B_R\left(\Sigma_X-\Sigma_X^{\rm eq}\frac{\Sigma_{\nu_R}}{\Sigma_{\nu_R}^{\rm eq}}\right),\\
\frac{d\Delta_X}{dy}&=\frac{\langle\Gamma_X\rangle}{s}\left[\Delta_X-\Sigma_X^{\rm eq}\left\{B_R\frac{\Delta_{\nu_R}(\Sigma_{e_R}^{\rm eq}+\Sigma_{\nu_R})}{\Sigma_{\nu_R}^{\rm eq}\Sigma_{e_R}^{\rm eq}}-B_L\frac{4(\Delta_X+\Delta_{\nu_R})}{\Sigma_L^{\rm eq}}-\epsilon\left(1-\frac{\Sigma_{\nu_R}}{\Sigma_{\nu_R}^{\rm eq}}\right)\right\}\right],\\
\frac{d\Delta_{\nu_R}}{dy}&=\frac{\langle\Gamma_X\rangle}{s}\left[-\epsilon\left(\Sigma_X-\Sigma_X^{\rm eq}\frac{\Sigma_{\nu_R}}{\Sigma_{\nu_R}^{\rm eq}}\right)-B_R\left(\Delta_X-\Sigma_X^{\rm eq}\frac{\Delta_{\nu_R}(\Sigma_{e_R}^{\rm eq}+\Sigma_{\nu_R})}{\Sigma_{\nu_R}^{\rm eq}\Sigma_{e_R}^{\rm eq}}\right)\right],
\label{eq:thermal_boltzmann_review}
\end{align}
Here $T$ is the temperature of the SM bath, $H$ is the Hubble rate,
$x=M_X/T$, and $y$ is defined by
$d/dy\equiv 3H(ds/dx)^{-1}d/dx$. Since $ds/dx<0$, $y$
decreases as the Universe cools and $x$ increases; the signs of the collision
terms in Eq.~\eqref{eq:thermal_boltzmann_review} follow this convention.
The quantities
$\langle\sigma v\rangle_{X\bar X}$ and $\langle\Gamma_X\rangle$ are the
thermally averaged annihilation cross section and decay rate.  The equilibrium
abundances $\Sigma_{e_R}^{\rm eq}$ and $\Sigma_L^{\rm eq}$ include particles
and antiparticles, summed over the three flavors and, for $L$, over weak
isospin.  The CP-even SM lepton densities are fixed at their equilibrium
values, while their asymmetries are eliminated using hypercharge and
total-lepton-number conservation.  The $B_R$ terms exchange asymmetry between
$X$ and $\nu_R$, the $B_L$ term accounts for the left-handed bath, and the
$\epsilon$ terms are the CP-odd sources. The factor $1/2$ in the
annihilation term follows from summing the equations for the two charge states:
in the CP-even limit, $Y_X=Y_{\bar X}=\Sigma_X/2$.

For the thermally averaged $X\bar X$ annihilation cross section we use the
hypercharge-mediated approximation of Ref.~\cite{Heeck:2023soj}.

The full analysis of Ref.~\cite{Heeck:2023soj} also includes $X$-mediated
$s$- and $t$-channel scatterings. We omit these processes and work in the
decay and inverse-decay approximation. To check its validity, we use
Eq.~(13) and the cross sections in Appendix~C of
Ref.~\cite{Heeck:2023soj}.  Here and below, $H(M_X)$ denotes the
radiation-dominated Hubble rate evaluated at $T=M_X$ in the absence of PBHs.
We define
\begin{equation}
  r_{s,t}^{\rm max}\equiv
  \max_x\left[
    \frac{\Gamma_{s,t}(x)}
    {H_R(x)+\Gamma_{\rm ID}(x)}
  \right],
  \label{eq:scattering_diagnostic_ratio}
\end{equation}
where $H_R$ is the radiation-only Hubble rate and $\Gamma_{\rm ID}$ is the
inverse-decay relaxation rate. We obtain $\Gamma_{s,t}$ from the
equilibrium densities and the thermally averaged cross sections in
Eqs.~(C5) and (C6) of Ref.~\cite{Heeck:2023soj}, using the on-shell subtraction
of Eq.~(C7), the coupling normalization of Eq.~(C8), and the same flavor and
particle-antiparticle counting conventions as for the equilibrium abundances
above.
In the PBH-assisted region with
$\Gamma_X/H(M_X)\lesssim10^{-2}$, which drives the extension to
$M_X\sim10^6$--$10^8\,{\rm GeV}$, we find
$r_{s,t}^{\rm max}\lesssim5\times10^{-2}$. At the two benchmarks shown in
Fig.~\ref{fig:etaB_evolution_benchmarks}, the ratio is approximately
$10^{-4}$ for $(M_X,\Gamma_X/H)=(10^7\,{\rm GeV},10^{-5})$ and
$2\times10^{-2}$ for $(10^{10}\,{\rm GeV},1)$. The scattering contribution
becomes noticeable only near the upper-right edge of the displayed plane,
which also motivates stopping the scan at $\Gamma_X/H(M_X)=10^4$.
In this comparison we use only the radiation contribution to the Hubble rate.  Including the PBH energy density would increase $H$ and make the scattering contribution even less important.  We also neglect the ordinary Higgs-neutrino Dirac Yukawa interactions, whose couplings are too small to equilibrate $\nu_R$ during the epoch considered here.

\section{PBH formation and evaporation} \label{sec:pbh_review}
Primordial black holes can form in the early Universe when sufficiently large overdensities re-enter the Hubble horizon and undergo gravitational collapse~\cite{Carr:1974nx,Carr:1975qj,Press:1973iz,Villanueva-Domingo:2021spv}.  Expressed in terms of primordial curvature fluctuations, this condition becomes a threshold criterion for the smoothed density contrast; the numerical threshold is sensitive to the perturbation shape and the background equation of state~\cite{Harada:2013epa,Musco:2018rwt,Carr:2020gox}.  We do not specify the microscopic origin of the enhanced primordial perturbations.  Instead, we parameterize the PBH population by its initial mass and abundance, which are the only ingredients needed for the evaporation-driven leptogenesis calculation below.

We assume a monochromatic population of neutral, non-rotating Schwarzschild PBHs~\cite{Sasaki:2018dmp,Carr:2020gox,Cheek:2021odj}\footnote{In Majorana leptogenesis, superradiance around rotating PBHs can enhance scalar production and source heavy right handed neutrinos~\cite{Ghoshal:2023fno}.}.  Extended mass functions, critical-collapse broadening, spin, charge, mergers and accretion are not included.  For the ultralight PBHs considered below, cosmological accretion is negligible compared with Hawking mass loss.
The monochromatic approximation isolates the dependence on a single evaporation time and makes the $(M_{\rm PBH},\beta')$ interpretation transparent.\footnote{Extended mass and spin distributions can cause different parts of the PBH population to evaporate at different times, modifying entropy injection, particle yields and $\Delta N_{\rm eff}$~\cite{Cheek:2022massspin}.  We do not explore this possibility here.}

We use the standard semiclassical Hawking law throughout the evaporation.  Proposed memory-burden effects can suppress late-time emission and alter the competition between nonthermal production and entropy dilution~\cite{Calabrese:2025sfh}; such modifications are not included here.

We denote cosmic time by $t$, the cosmological scale factor by $a$ and the
Hubble rate by $H=\dot a/a$, where the dot denotes a derivative with respect
to $t$.  PBH formation occurs at $t_{\rm f}$, with
$T_{\rm f}=T(t_{\rm f})$ and $a_{\rm f}=a(t_{\rm f})$.  The physical energy densities of the SM radiation bath and the
PBH population are $\rho_R$ and $\rho_{\rm PBH}$, while $\rho_{\rm tot}$
denotes the sum over all components.  Throughout the manuscript,
$M_{\rm PBH}$ denotes the initial PBH mass, while $M=M(t)\equiv M[a(t)]$ is the
instantaneous mass during evaporation, with $M(t_{\rm f})=M_{\rm PBH}$.  In particular,
$T_{\rm BH}$ and all greybody emission functions depend on this instantaneous mass rather than
on the initial mass once evaporation begins.

During radiation domination the horizon mass at formation is
\begin{equation}
  M_H(T_{\rm f})
  =\frac{4\pi}{3}\rho_R(T_{\rm f})H_{\rm f}^{-3}
  =\frac{1}{2GH_{\rm f}},
\end{equation}
where $H_{\rm f}\equiv H(T_{\rm f})$.  We write the initial PBH mass as
\begin{equation}
  M_{\rm PBH}=\gamma M_H(T_{\rm f}),
\end{equation}
where $\gamma\simeq0.2$ parametrizes the collapse efficiency~\cite{Carr:2009jm,Carr:2020gox}.  Using $\rho_R=(\pi^2/30)g_\star(T_{\rm f})T_{\rm f}^4$ and $H_{\rm f}^2=8\pi G\rho_R(T_{\rm f})/3$, the PBH formation temperature is
\begin{align}
  T_{\rm f}
  & =
  \left(\frac{45\gamma^2}{16\pi^3 g_\star(T_{\rm f})}\right)^{1/4}
  \left(\frac{M_{\rm Pl}}{M_{\rm PBH}}\right)^{1/2}M_{\rm Pl}
  \nonumber\\
  & \simeq
  4.3\times10^{15}\,{\rm GeV}
  \left(\frac{\gamma}{0.2}\right)^{1/2}
  \left(\frac{106.75}{g_\star(T_{\rm f})}\right)^{1/4}
  \left(\frac{1\,{\rm g}}{M_{\rm PBH}}\right)^{1/2}.
  \label{eq:pbh_formation_temperature}
\end{align}
Here $G$ is Newton's constant and $M_{\rm Pl}\equiv G^{-1/2}$ is the
non-reduced Planck mass.
When gram masses are inserted in the evolution equations, we use
$1\,{\rm g}=5.6096\times10^{23}\,{\rm GeV}$ and
$M_{\rm Pl}=1.2209\times10^{19}\,{\rm GeV}$.  These conventions give the
numerical coefficient in Eq.~\eqref{eq:pbh_formation_temperature}.
  The functions $g_\star$ and $g_{\star s}$ denote the
effective relativistic degrees of freedom contributing to the energy and
entropy densities of the thermal bath, respectively.
In the numerical analysis we set $g_\star(T_{\rm f})=106.75$ and neglect the small contribution from the additional fields.

We denote the initial PBH energy fraction by
\begin{equation}
  \beta\equiv\left.\frac{\rho_{\rm PBH}}{\rho_{\rm tot}}\right|_{T_{\rm f}}.
\end{equation}
More generally, we denote the evolving PBH energy fraction by
$\Omega_{\rm PBH}(a)\equiv\rho_{\rm PBH}(a)/\rho_{\rm tot}(a)$, so that
$\beta=\Omega_{\rm PBH}(a_{\rm f})$ at formation.  For $\beta\ll1$ this agrees
with the common convention $\rho_{\rm PBH}/\rho_R$ at formation.  We also use
the rescaled abundance
\begin{equation}
  \beta'
  =
  \gamma^{1/2}
  \left(\frac{g_\star(T_{\rm f})}{106.75}\right)^{-1/4}
  \beta,
  \label{eq:beta_prime_def}
\end{equation}
which absorbs the conventional dependence on the collapse efficiency and the relativistic degrees of freedom at formation.

Once formed, a Schwarzschild PBH evaporates by Hawking emission~\cite{Hawking:1974rv,Hawking:1975vcx,Carr:1976zz,Page:1976df,Page:1977um}.  Its instantaneous temperature is
\begin{equation}
  T_{\rm BH}(M)=\frac{1}{8\pi G M}
  \simeq
  1.06\times10^7\,{\rm GeV}
  \left(\frac{10^6\,{\rm g}}{M}\right)
  =1.06\times10^{13}\,{\rm GeV}
  \left(\frac{1\,{\rm g}}{M}\right).
  \label{eq:pbh_temperature}
\end{equation}
The inverse relation $T_{\rm BH}\propto M^{-1}$ is central to PBH-assisted
leptogenesis: a heavy state of mass $m_i$ is efficiently emitted once
$T_{\rm BH}$ approaches $m_i$, or equivalently
\begin{equation}
  M\lesssim
  1.06\times10^{6}\,{\rm g}
  \left(\frac{10^7\,{\rm GeV}}{m_i}\right).
  \label{eq:pbh_threshold_mass}
\end{equation}
For $T_{\rm BH}<m_i$, particle emission remains possible through the
high-energy tail of the Hawking spectrum, but is exponentially suppressed.  A
PBH that is initially colder than $m_i$ can therefore emit the particle during
its final, hotter stage.  For the benchmark masses
$M_{\rm PBH}=10^{-1}$--$10^2\,{\rm g}$, the initial Hawking temperature is
approximately $10^{11}$--$10^{14}\,{\rm GeV}$.

For a particle species $i$ with spin $s_i$, mass $m_i$, internal degrees of freedom $g_i$, momentum $p$ and energy $E_i=\sqrt{p^2+m_i^2}$, the differential emission spectrum is~\cite{Page:1976df,Page:1977um,MacGibbon:1990zk,MacGibbon:1991tj,Cheek:2021odj}
\begin{equation}
  \frac{\dd^2 N_i}{\dd p\,\dd t}
  =
  \frac{g_i}{2\pi^2}
  \frac{\sigma_{s_i}(M,m_i,p)\,p^3/E_i}
  {\exp(E_i/T_{\rm BH})-(-1)^{2s_i}}.
  \label{eq:pbh_emission_rate}
\end{equation}
Here $N_i$ is the number of emitted particles and $\sigma_{s_i}$ is the
spin-dependent absorption cross section.  Equivalently, one may express the
spectrum in terms of dimensionless partial-wave greybody factors related to
$\sigma_{s_i}$ by the standard absorption formulae.  The emission is gravitational and is therefore independent of the Standard Model gauge charges
of the emitted particle, up to its spin, mass threshold and greybody factor.
In Sec.~\ref{sec:pbh_dirac} we use greybody-integrated number and power
functions based on the massive, spin-dependent fits of
Ref.~\cite{Cheek:2021odj}.  Their continuous dependence on $m_i/T_{\rm BH}$
retains the suppressed emission below threshold.

All PBHs in the displayed mass range evaporate while the visible bath is above
electroweak symmetry breaking.  The Hawking emission function includes the SM
particle content, while the two charge states of $X$ and the three $\nu_R$
flavors are added separately to avoid double counting.  The bath evolution
accounts for changes in $g_\star(T)$ and $g_{\star s}(T)$.

The same Hawking process determines the PBH mass-loss rate.  Integrating Eq.~\eqref{eq:pbh_emission_rate} with an additional factor of $E_i$ gives the power emitted into species $i$.  Summing over the species included in the numerical treatment gives
\begin{equation}
  \frac{\dd M}{\dd t}
  =
  -{\cal E}(M)\frac{M_{\rm Pl}^4}{M^2},
  \label{eq:pbh_mass_loss}
\end{equation}
where ${\cal E}(M)$ is the dimensionless, greybody-weighted total emission
function.  We include the SM species, both scalar charge states and all three right handed neutrino flavors.  For constant ${\cal E}$, the lifetime scales
as $\tau_{\rm PBH}\simeq M_{\rm PBH}^3/(3{\cal E}M_{\rm Pl}^4)$~\cite{MacGibbon:1990zk,MacGibbon:1991tj,Cheek:2021odj}.  In the
numerical calculation, we integrate Eq.~\eqref{eq:pbh_mass_loss} and terminate
the semiclassical evolution near $M=M_{\rm Pl}$.
For the smallest initial mass considered, the unevaporated fraction at this
cutoff is at most $M_{\rm Pl}/M_{\rm PBH}\simeq2.2\times10^{-4}$, so the
untracked endpoint affects the total injected energy only below the
per-mille level.

Prior to appreciable mass loss, the PBH component behaves as pressureless
matter, $\rho_{\rm PBH}\propto a^{-3}$, while the bath scales as
$\rho_R\propto a^{-4}$.  Their ratio therefore increases with the scale
factor.  We define the onset of PBH domination numerically as the initial
abundance for which the maximum value of $\Omega_{\rm PBH}$ during the
evolution reaches $1/2$.  We determine this boundary directly from the
numerical evolution.
For orientation, the radiation-dominated analytical estimate gives
\begin{equation}
  \beta'_c\simeq2.6\times10^{-6}
  \left(\frac{M_{\rm PBH}}{1\,{\rm g}}\right)^{-1},
  \label{eq:beta_prime_domination_estimate}
\end{equation}
for the conventions of Eqs.~\eqref{eq:pbh_formation_temperature} and~\eqref{eq:beta_prime_def}.  This estimate is useful for scaling arguments, but the numerical line shown in Fig.~\ref{fig:pbh_plane} is the criterion used in the analysis.  Its normalization can differ from analytic estimates in the literature because of the precise definition of $\beta'$, the collapse parameter $\gamma$, the greybody-weighted lifetime and the choice of evaporation temperature.
Similarly, in the radiation-dominated approximation, the evaporation temperature scales as
\begin{equation}
  T_{\rm ev}^{\rm RD}\sim
  \left(\frac{45}{16\pi^3g_\star(T_{\rm ev})}\right)^{1/4}
  \sqrt{\frac{M_{\rm Pl}}{\tau_{\rm PBH}}}
  \propto M_{\rm PBH}^{-3/2}.
  \label{eq:Tev_scaling_revision}
\end{equation}
Heavier PBHs therefore evaporate later, at a lower plasma temperature, when
thermal production of $X$ and inverse decays are less efficient. This scaling
helps interpret the mass dependence of the results, while the numerical
calculation follows the PBH mass loss and the Friedmann evolution directly.

The ultralight PBHs used in the scans evaporate long before Big Bang
nucleosynthesis (BBN), so we do not apply bounds from late evaporation during
or after nucleosynthesis.

\section{PBH Dirac Leptogenesis Boltzmann Equations} \label{sec:pbh_dirac}
Our numerical implementation builds on ULYSSES~\cite{Granelli:2020pim,Granelli:2023vcm,Granelli:2026goh}, which we extend to the Dirac leptogenesis system considered here.  We parameterize the rest frame decay width through
$\Gamma_X/H(M_X)$, using the convention defined in
Sec.~\ref{sec:dirac_review}.  This ratio fixes $\Gamma_X$, while the
Hubble rate entering the subsequent evolution includes all energy components,
as in coupled PBH Boltzmann treatments of Hawking
production~\cite{Cheek:2021odj,Cheek:2022dbx}.  Small values of this ratio
correspond to a thermal source that develops slowly compared with the
expansion, whereas larger values make decays and inverse decays more
efficient.  Hawking emission provides an additional production history for
$X$, with normalization and timing set by the PBH population.

As a check of the thermal limit, we switched off the PBH energy density and
all Hawking sources and verified that the calculation reproduces the
characteristic structure found in Ref.~\cite{Heeck:2023soj}, including the
low-rate boundary and the region in which the asymmetry changes sign.  This
comparison is made where the omitted $s$- and $t$-channel scatterings are
negligible according to the diagnostic in
Eq.~\eqref{eq:scattering_diagnostic_ratio}.

The notation used below differs slightly from that of
Sec.~\ref{sec:dirac_review}.  The quantity $M_{\rm PBH}$ denotes the initial
PBH mass, while $M(a)$ is its instantaneous value during evaporation.  That
section uses entropy-normalized yields $Y_A$, whereas here $\Sigma_A$ and
$\Delta_A$ denote the fixed-reference comoving abundances defined below.
Superscripts $\Therm$ and $\BH$ distinguish thermally produced and
PBH-produced populations.
With this notation, we combine Secs.~\ref{sec:dirac_review} and~\ref{sec:pbh_review} using the scale
factor $a$ as the independent variable.  At the beginning of the integration,
when the SM bath has temperature $T_i$, we set $a_i=1$ and define
\begin{equation}
  n_{\rm ref}\equiv n_\gamma(T_i)=\frac{2\zeta(3)}{\pi^2}T_i^3,
  \qquad \Sigma_A\equiv\frac{a^3(n_A+n_{\bar A})}{n_{\rm ref}},
  \qquad \Delta_A\equiv\frac{a^3(n_A-n_{\bar A})}{n_{\rm ref}}.
  \label{eq:pbh_reference_density}
\end{equation}
Since $n_{\rm ref}$ is fixed at the beginning of the integration, this
normalization keeps the entropy injected by PBH evaporation explicit. All
equilibrium abundances below use the same normalization.

Following Ref.~\cite{Heeck:2023soj}, we initialize the thermal mediator and
right handed neutrino abundances to zero.  We also set
$\Sigma_X^{\BH}=\Delta_X=\Delta_{\nu_R}=R_X^{\BH}=R_{\nu_R}=0$.  This choice
describes a thermal SM bath with no primordial population or asymmetry in the
additional particle sectors.  It does not prevent the thermal mediator from
approaching equilibrium, since gauge-mediated $X\bar X$ production and inverse
decays are included in its Boltzmann equation.

For the PBH runs shown below, the coupled evolution begins at PBH formation,
with $T_i=T_{\rm f}$, $M=M_{\rm PBH}$ and
$\rho_{\rm PBH}/\rho_{\rm tot}=\beta$; the initial radiation density is
$\rho_R(T_i)=(\pi^2/30)g_\star(T_i)T_i^4$.  The no-PBH evolution starts at
$T_i=100M_X$.  If this thermal starting temperature lies above
$T_{\rm f}$, we first evolve the thermal system to PBH formation and use the
resulting abundances as initial data for the coupled PBH evolution.

We define the comoving energy variables through $\rho_R=\varrho_Ra^{-4}$ and $\rho_{\PBH}=\varrho_{\PBH}a^{-3}$.  Thus $\varrho_R$ changes when entropy is injected and $\varrho_{\PBH}$ changes when the PBH mass decreases.  Hawking emission adds the PBH-produced component $\Sigma_X^{\BH}$, so we split $\Sigma_X=\Sigma_X^{\Therm}+\Sigma_X^{\BH}$.  We evolve the total mediator asymmetry $\Delta_X$.  Since a neutral Schwarzschild PBH emits $X$ and $\bar X$ at equal rates, the PBH-produced component has no independent asymmetry source.  For $\Delta_X^{\BH}(a_i)=0$, it remains zero, and therefore $\Delta_X=\Delta_X^{\Therm}$.

To avoid confusing the Hawking emission functions with the Yukawa matrix
$F_i$ in Eq.~\eqref{eq:dirac_lagrangian_review}, we denote the integrated number emission rate by
$\Gamma_{\PBH\to i}(M)$ and the emitted power by $\mathcal P_i(M)$.  The
corresponding dimensionless emission function is $\mathcal F_i(M)$:

\begin{equation}
  \Gamma_{\PBH\to i}=\int_0^\infty dp\,\frac{d^2N_i}{dp\,dt},
  \qquad
  \mathcal P_i=\int_0^\infty dp\,E_i\frac{d^2N_i}{dp\,dt},
  \qquad
  \mathcal F_i\equiv G^2M^2\mathcal P_i.
\end{equation}
For the particle content retained here, the total emission function is
$\mathcal F_T=\mathcal F_{\SM}+\mathcal F_X+\mathcal F_{\nu_R}={\cal E}(M)$.
Here the three terms are the contributions from the SM bath, the scalar sector
and the right handed neutrino sector.  The scalar term includes both charge
states, while the neutrino term includes neutrinos and antineutrinos of all
three flavors~\cite{Cheek:2021odj}.  The background obeys
\begin{align}
 aH\frac{dM}{da}&=-\frac{\mathcal F_T}{G^2M^2},
 &aH\frac{d\varrho_{\PBH}}{da}&=\frac{\dot M}{M}\varrho_{\PBH},
 \label{eq:bg_M}\\
 aH\frac{d\varrho_R}{da}&=-\frac{\mathcal F_{\SM}}{\mathcal F_T}\frac{\dot M}{M}a\varrho_{\PBH}+\mathcal C_R,
 \label{eq:bg_rhoR}
\end{align}
where a dot denotes a derivative with respect to cosmic time and
$\dot M=-\mathcal F_T/(G^2M^2)$.  The term $\mathcal C_R$ collects the energy
exchanged between the SM radiation bath and the $X$ and $\nu_R$ sectors.  Its
explicit form is given in Eq.~\eqref{eq:radiation_collision_operator}, after
the corresponding abundance and energy equations have been introduced.
We write the thermal mediator energy density as
$\rho_X^{\Therm}=n_X^{\Therm}\langle E_X\rangle_{\Therm}$, where
$n_X^{\Therm}$ counts both $X$ and $\bar X$, and use
$\langle E_X\rangle_{\Therm}=M_XK_1(x)/K_2(x)+3T$, with $x=M_X/T$ and
$K_1$ and $K_2$ the modified Bessel functions of the second kind. The quantities $\rho_X^{\BH}$
and $\rho_{\nu_R}$ are the energy densities of the PBH-produced $X+\bar X$
population and of right handed neutrinos and antineutrinos summed over the
three flavors.  The Friedmann equation is then
\begin{equation}
 H^2=\frac{8\pi G}{3}\rho_{\rm tot},
 \qquad
 \rho_{\rm tot}=\rho_R+\rho_{\PBH}+\rho_X^{\Therm}
 +\rho_X^{\BH}+\rho_{\nu_R}.
 \label{eq:full_hubble}
\end{equation}
The temperature follows from $\rho_R=(\pi^2/30)g_\star(T)T^4$ and
$s=(2\pi^2/45)g_{\star s}(T)T^3$:
\begin{equation}
 aH\frac{dT}{da}=-\frac{T}{\Delta_T}\left[H-\frac{g_\star}{g_{\star s}}\frac{aH}{4\varrho_R}\frac{d\varrho_R}{da}\right],
 \qquad
 \Delta_T=1+\frac{T}{3g_{\star s}}\frac{dg_{\star s}}{dT}.
 \label{eq:bg_temperature}
\end{equation}
Here $\Delta_T$ accounts for the temperature dependence of the entropy degrees
of freedom.

With the normalization in Eq.~\eqref{eq:pbh_reference_density}, the Hawking
source entering the comoving abundance of species $i$ is~\cite{Cheek:2021odj}
\begin{equation}
  \mathcal S_i^{\PBH}=a^3\frac{\Gamma_{\PBH\to i}n_{\PBH}}{n_{\rm ref}},
  \qquad n_{\PBH}=\frac{\rho_{\PBH}}{M}.
  \label{eq:pbh_number_source}
\end{equation}
Here $n_{\PBH}$ is the physical PBH number density.  The factor $a^3$ follows from the comoving normalization.  The rate $\Gamma_{\PBH\to X}$ includes
both $X$ and $\bar X$, while $\Gamma_{\PBH\to\nu_R}$ includes the three right
handed neutrino flavors and their antiparticles.  The same state sums are used
in $\mathcal F_X$ and $\mathcal F_{\nu_R}$.
The Hawking integrals retain the quantum-statistical denominator in
Eq.~\eqref{eq:pbh_emission_rate}.  Plasma collision terms follow the
momentum-averaged Maxwell-Boltzmann treatment of
Ref.~\cite{Heeck:2023soj}, without final-state Bose enhancement or Pauli
blocking.
 The $X$ particles emitted by PBHs
are not in thermal equilibrium, so their mean energy cannot be inferred from
the plasma temperature. We instead track their number and energy densities
separately and define
\begin{equation}
  R_X^{\BH}=\frac{a^4\rho_X^{\BH}}{\rho_{R,i}},
  \qquad
  \langle E_X\rangle_{\BH}=\frac{\rho_X^{\BH}}{n_X^{\BH}},
  \qquad
  w_X^{\BH}=\frac13\left[1-\left(\frac{M_X}{\langle E_X\rangle_{\BH}}\right)^2\right].
  \label{eq:xbh_energy_moment}
\end{equation}
Here $\rho_{R,i}=\rho_R(T_i)$, $R_X^{\BH}$ is a dimensionless
radiation-scaled energy density, and $n_X^{\BH}$ is the physical number
density of PBH-produced $X$ and $\bar X$. The last relation approximates the
effective equation-of-state parameter
$w_X^{\BH}\equiv p_X^{\BH}/\rho_X^{\BH}$ using the mean energy of this
population. It approaches $1/3$ in the relativistic limit and zero in the
nonrelativistic limit, and determines the redshifting term in
Eq.~\eqref{eq:RXBH_evolution}. The mean energy also sets the effective decay
rates,
\begin{equation}
  \Gamma_X^{\BH}=\Gamma_X\frac{M_X}{\langle E_X\rangle_{\BH}},
  \qquad
  \Gamma_X^{\Therm}=\Gamma_X\frac{K_1(x)}{K_2(x)}
  \equiv\langle\Gamma_X\rangle.
\end{equation}
Relativistic $X$ particles decay more slowly than particles at rest. For the
PBH-produced population this time dilation is described by
$M_X/\langle E_X\rangle_{\BH}$, while $K_1(x)/K_2(x)$ is the corresponding
thermal average. The abundance equations are
\begin{align}
  aH\frac{d\Sigma_X^{\Therm}}{da}&=\frac12\langle\sigma v\rangle_{X\bar X}n_{\rm ref}a^{-3}\left[(\Sigma_X^{\eq})^2-(\Sigma_X^{\Therm})^2\right]
  -\Gamma_X^{\Therm}\left[\Sigma_X^{\Therm}-\Sigma_X^{\eq}\left(B_L+B_R\frac{\Sigma_{\nu_R}}{\Sigma_{\nu_R}^{\eq}}\right)\right],\\
  aH\frac{d\Sigma_X^{\BH}}{da}&=-\Gamma_X^{\BH}\Sigma_X^{\BH}+\mathcal S_X^{\PBH},\\
  aH\frac{d\Sigma_{\nu_R}}{da}&=B_R\Gamma_X^{\Therm}\left(\Sigma_X^{\Therm}-\Sigma_X^{\eq}\frac{\Sigma_{\nu_R}}{\Sigma_{\nu_R}^{\eq}}\right)+B_R\Gamma_X^{\BH}\Sigma_X^{\BH}+\mathcal S_{\nu_R}^{\PBH}.
  \label{eq:pbh_even_system}
\end{align}
The first equation tracks the thermal $X$ abundance. Annihilations drive it
towards equilibrium, while decays and inverse decays exchange particles with
the left and right handed sectors. The second equation balances Hawking
production of PBH-produced $X$ against their decays, while the third contains
the three sources of $\nu_R+\bar\nu_R$: decays of thermally produced $X$,
decays of PBH-produced $X$ and direct Hawking emission.
The PBH-produced population enters Eqs.~\eqref{eq:pbh_even_system}
and~\eqref{eq:pbh_odd_system} through Hawking production and its subsequent
decays.  Schwarzschild PBHs emit $X$ and $\bar X$ symmetrically, so
$\Delta_X^{\BH}=0$, while inverse decays populate the thermal component \footnote{The PBH-produced population is evolved only through its number
and mean-energy moments.  We do not include its elastic scattering with the
plasma or its annihilations with PBH-produced and thermally produced
antiparticles.  Evaluating these collision terms requires an additional
prescription for the evolving nonthermal phase-space distribution and is left
for future work.}.

The corresponding asymmetry equations are
\begin{align}
  aH\frac{d\Delta_X}{da}&=-\Gamma_X^{\Therm}\left[\Delta_X-\Sigma_X^{\eq}\left(B_R\frac{\Delta_{\nu_R}(\Sigma_{e_R}^{\eq}+\Sigma_{\nu_R})}{\Sigma_{\nu_R}^{\eq}\Sigma_{e_R}^{\eq}}-B_L\frac{4(\Delta_X+\Delta_{\nu_R})}{\Sigma_L^{\eq}}-\epsilon\left[1-\frac{\Sigma_{\nu_R}}{\Sigma_{\nu_R}^{\eq}}\right]\right)\right],\\
  aH\frac{d\Delta_{\nu_R}}{da}&=\epsilon\Gamma_X^{\Therm}\left(\Sigma_X^{\Therm}-\Sigma_X^{\eq}\frac{\Sigma_{\nu_R}}{\Sigma_{\nu_R}^{\eq}}\right)+\epsilon\Gamma_X^{\BH}\Sigma_X^{\BH}
  +B_R\Gamma_X^{\Therm}\left[\Delta_X-\Sigma_X^{\eq}\frac{\Delta_{\nu_R}}{\Sigma_{\nu_R}^{\eq}}\left(1+\frac{\Sigma_{\nu_R}}{\Sigma_{e_R}^{\eq}}\right)\right].
  \label{eq:pbh_odd_system}
\end{align}
In the equation for $\Delta_{\nu_R}$, the first term proportional to
$\epsilon$ is the thermal CP-odd source, generated by decays and inverse
decays of the thermally produced $X$ population. The second,
$\epsilon\Gamma_X^{\BH}\Sigma_X^{\BH}$, is the corresponding source from
decays of the $X$ population emitted by PBHs. The remaining terms transfer or
wash out an existing asymmetry. Direct Hawking emission of $\nu_R$ does not
appear because it produces $\nu_R$ and $\bar\nu_R$ at equal rates.

The energy moment of the PBH-produced mediator population obeys
\begin{equation}
  aH\frac{dR_X^{\BH}}{da}=\frac{a\varrho_{\PBH}}{\rho_{R,i}}\frac{\mathcal F_X}{\mathcal F_T}\left(-\frac{\dot M}{M}\right)-\Gamma_X^{\BH}R_X^{\BH}+H(1-3w_X^{\BH})R_X^{\BH}.
  \label{eq:RXBH_evolution}
\end{equation}
The three terms on the right hand side describe Hawking energy injection,
energy lost through $X$ decays and the redshifting of the PBH-produced
population. Since $R_X^{\BH}$ contains a factor $a^4$, the last term vanishes
for a relativistic population, with $w_X^{\BH}=1/3$, and accounts for the
slower redshift as the particles become nonrelativistic.
The right handed neutrino
radiation moment $R_{\nu_R}=a^4\rho_{\nu_R}/\rho_{R,i}$ satisfies
\begin{equation}
  \scalebox{0.95}{$\displaystyle
  aH\frac{dR_{\nu_R}}{da}
  =\frac{a n_{\rm ref}}{\rho_{R,i}}\left[\frac{\langle E_X\rangle_{\Therm}}{2}B_R\Gamma_X^{\Therm}\Sigma_X^{\Therm}+\frac{M_X}{2}B_R\Gamma_X\Sigma_X^{\BH}\right]
  -B_R\Gamma_X^{\Therm}\frac{\Sigma_X^{\eq}}{\Sigma_{\nu_R}^{\eq}}R_{\nu_R}
  +\frac{a\varrho_{\PBH}}{\rho_{R,i}}\frac{\mathcal F_{\nu_R}}{\mathcal F_T}\left(-\frac{\dot M}{M}\right).
  $}
  \label{eq:RnuR}
\end{equation}
The two terms in square brackets describe decays of thermal and PBH-produced
mediators, respectively. The negative term accounts for
the reverse transfer through inverse decays, while the last term describes
direct Hawking injection of right handed neutrinos. The prefactor
$a n_{\rm ref}/\rho_{R,i}$ converts decay terms written in terms of comoving
abundances into the radiation moment $R_{\nu_R}$. In each $X$ decay the right
handed neutrino receives, on average, half of the parent energy. For
PBH-produced $X$, combining this energy with the time-dilated decay rate gives
the factor $\Gamma_XM_X/2$ in Eq.~\eqref{eq:RnuR}.
Equation~\eqref{eq:pbh_even_system} evolves the total number abundance of
$\nu_R+\bar\nu_R$, while Eq.~\eqref{eq:RnuR} evolves their energy density.
We do not resolve their full momentum distribution, so the inverse-decay terms
are approximate when the nonthermal component becomes important.

With the abundance and energy equations now specified, the collision term in
Eq.~\eqref{eq:bg_rhoR} can be written as
\begin{equation}
  \mathcal C_R=-\mathcal Q_X^{\Therm}-\mathcal Q_{\nu_R}^{\Therm}
  +\left(1-\frac{B_R}{2}\right)
   a n_{\rm ref}\Gamma_XM_X\Sigma_X^{\BH},
  \label{eq:radiation_collision_operator}
\end{equation}
where
\begin{align}
 \mathcal Q_X^{\Therm}={}&a n_{\rm ref}\langle E_X\rangle_{\Therm}
 \bigg\{
 \frac{1}{2}\langle\sigma v\rangle_{X\bar X}n_{\rm ref}a^{-3}
 \left[(\Sigma_X^{\eq})^2-(\Sigma_X^{\Therm})^2\right]
 \nonumber\\[-1mm]
 &\hspace{34mm}
 -\Gamma_X^{\Therm}\left[
 \Sigma_X^{\Therm}-\Sigma_X^{\eq}\left(
 B_L+B_R\frac{\Sigma_{\nu_R}}{\Sigma_{\nu_R}^{\eq}}
 \right)\right]\bigg\},
 \label{eq:thermal_X_energy_transfer}\\
 \mathcal Q_{\nu_R}^{\Therm}={}&
 \frac{a n_{\rm ref}\langle E_X\rangle_{\Therm}}{2}
 B_R\Gamma_X^{\Therm}\Sigma_X^{\Therm}
 -B_R\Gamma_X^{\Therm}
 \frac{\Sigma_X^{\eq}}{\Sigma_{\nu_R}^{\eq}}
 a^4\rho_{\nu_R}.
 \label{eq:thermal_nuR_energy_transfer}
\end{align}
Here $\mathcal Q_X^{\Therm}$ is the collision-induced change in the comoving
energy of the thermal $X+\bar X$ population, evaluated with
$\rho_X^{\Therm}=n_X^{\Therm}\langle E_X\rangle_{\Therm}$, and
$\mathcal Q_{\nu_R}^{\Therm}$ is the net energy transferred to
$\nu_R+\bar\nu_R$ by thermal decays and inverse decays.  A positive
$\mathcal Q$ denotes energy gained by that sector, which fixes the two minus
signs in Eq.~\eqref{eq:radiation_collision_operator}.  The factor $1/2$ in
the annihilation term accounts for the use of an abundance containing both
charge states.  In the massless two-body approximation, the right handed
neutrino receives on average one half of the parent energy.  The visible
fraction of a PBH-produced $X$ decay is therefore
$B_L+B_R/2=1-B_R/2$.  Multiplication by the time-dilated rate
$\Gamma_XM_X/\langle E_X\rangle_{\BH}$ cancels the mean parent energy and
gives the last term in Eq.~\eqref{eq:radiation_collision_operator}.
  Direct Hawking injection
into the SM bath is already the first term in Eq.~\eqref{eq:bg_rhoR}, while
the matching injections into the $X$ and $\nu_R$ sectors appear in
Eqs.~\eqref{eq:RXBH_evolution} and~\eqref{eq:RnuR}.

Together, these equations form the closed energy-transfer system used in the
numerical evolution.  When the component equations are summed, transfers
between the PBHs, the two $X$ populations, the SM bath and the $\nu_R$ sector
cancel, as in greybody-integrated PBH Boltzmann systems
~\cite{Cheek:2021odj,Cheek:2022dbx}.  We check this cancellation in every run
by reconstructing the total energy density and pressure and evaluating the
integrated residual of
\begin{equation}
  \dot\rho_{\rm tot}+3H\left(\rho_{\rm tot}+p_{\rm tot}\right)=0.
  \label{eq:energy_conservation_check_revision}
\end{equation}
Here $p_{\rm tot}$ is the total pressure summed over all components.  The
condition is well satisfied throughout the numerical evolution.\footnote{As
a representative example, for $M_X=10^8\,{\rm GeV}$,
$\Gamma_X/H(M_X)=10^{-4}$ and $\beta'=10^{-4}$, the accumulated relative
residual never exceeds $4\times10^{-5}$ without PBHs and remains below
$3\times10^{-6}$ for $M_{\rm PBH}=10^{-1},10^1$ and $10^2\,{\rm g}$.}

To estimate the uncertainty associated with using a mean energy for the
PBH-produced population, we repeated the calculation after increasing
$\Gamma_X^{\BH}$ by factors of two and three, while always requiring
$\Gamma_X^{\BH}\leq\Gamma_X$. The final asymmetry changes by less than about
$0.4\%$ over most of the parameter space and by up to about $10\%$ only for
the smallest masses and decay rates considered. The final asymmetry is
therefore stable under moderate variations of the effective decay rate of the
PBH-produced mediator.

We model sphaleron freeze-out as an instantaneous transition at
$T_{\rm sph}=131.7\,{\rm GeV}$. Below this temperature, we fix the comoving
baryon number while continuing the background evolution to account for any
subsequent entropy injection.  The final baryon-to-photon
ratio is evaluated as
\begin{equation}
  \etaB=
  \frac{28}{79}
  \frac{\gstarS(T_0)}{\gstarS(T_{\rm end})}
  \frac{n_{\rm ref}\Delta_{\nu_R}(T_{\rm sph})}
  {a_{\rm end}^3n_\gamma(T_{\rm end})},
  \label{eq:etaB_observable}
\end{equation}
where $n_\gamma(T)=2\zeta(3)T^3/\pi^2$, $T_{\rm end}$ is the final integration
temperature, $a_{\rm end}=a(T_{\rm end})$, and $T_0$ is a late reference
temperature after electron-positron annihilation.  For points that produce at
least the observed asymmetry at the reference value of $\epsilon$,
we find $|\Delta_X(T_{\rm sph})|\ll|\Delta_{\nu_R}(T_{\rm sph})|$.  The
visible-sector asymmetry can therefore be determined from
$\Delta_{\nu_R}(T_{\rm sph})$ alone.
As a normalization check, combining the comoving variables in
Eq.~\eqref{eq:pbh_reference_density} with the sphaleron conversion in
Eq.~\eqref{eq:sphaleron_conversion_review}, the no-PBH limit of
Eq.~\eqref{eq:etaB_observable} reduces, for conserved comoving entropy, to the
standard relation
$\eta_B\simeq7.04\,Y_B=7.04(28/79)Y_{\Delta_{\nu_R}}(T_{\rm sph})$.
The additional scale-factor and $g_{\star s}$ factors in
Eq.~\eqref{eq:etaB_observable} account for entropy injection during the
subsequent evolution.
The same evolution also determines the energy stored in the right handed
neutrino sector and hence its contribution to dark radiation.  Following the
energy-density treatment of Ref.~\cite{Cheek:2022dbx}, we compute
\begin{equation}
  \Delta N_{\rm eff}
  =
  \left[
    \frac{8}{7}\left(\frac{4}{11}\right)^{-4/3}
    +N_{\rm eff}^{\rm SM}
  \right]
  \frac{\rho_{\nu_R}(T_{\rm end})}{\rho_R(T_{\rm end})}
  \frac{g_\star(T_{\rm end})}{g_\star(T_{\rm eq})}
  \left[
    \frac{g_{\star s}(T_{\rm eq})}{g_{\star s}(T_{\rm end})}
  \right]^{4/3},
  \label{eq:DNeff_definition}
\end{equation}
where $T_{\rm eq}=0.75\,{\rm eV}$ is the reference temperature used for the
late-time conversion and $N_{\rm eff}^{\rm SM}=3.045$~\cite{deSalas:2016ztq,Planck:2018vyg}.
This includes both directly emitted $\nu_R$ and neutrinos produced in mediator
decays.
The two observables retain complementary information from the same cosmological history. The baryon asymmetry is determined by the
signed value of $\Delta_{\nu_R}$ at sphaleron freeze-out, followed by any
subsequent entropy dilution. By contrast, $\Delta N_{\rm eff}$ depends on the
total $\nu_R+\bar\nu_R$ energy density at the end of the evolution, including
symmetric production from Hawking emission. This distinction will be useful
when interpreting the PBH parameter scan in Fig.~\ref{fig:pbh_plane}.

\section{Results} \label{sec:results}

We organize the numerical results by first following the Boltzmann evolution
in Fig.~\ref{fig:boltzmann_evolution} and then moving to the parameter scans.
This allows us to identify how the
PBH sources modify the different particle populations before discussing their
impact on the final baryon asymmetry. We then vary the PBH initial mass and
abundance in Fig.~\ref{fig:pbh_plane} and compare the baryon asymmetry with the
associated dark radiation signal before mapping the regions with sufficient
asymmetry in the particle parameter plane in
Fig.~\ref{fig:particle_plane_etaB}. The two
representative evolutions in Fig.~\ref{fig:etaB_evolution_benchmarks} show
explicitly why PBHs can enhance the asymmetry in one region and reduce it in
another.
The main analysis uses $B_R=0.99$, $B_L=0.01$ and $\epsilon=10^{-4}$.  We
briefly discuss below how the complementary choice $B_R=0.01$, $B_L=0.99$
affects the baryon asymmetry and dark radiation, with additional results collected in
Appendix~\ref{app:branching_ratio}.  For all figures in which the initial PBH
abundance is fixed, we use $\beta'=10^{-4}$.
As shown below by the numerical PBH domination boundary in
Fig.~\ref{fig:pbh_plane}, this value lies above the domination threshold for
all the PBH masses considered.  It therefore selects a regime in which PBH
evaporation can appreciably affect Dirac leptogenesis.  For smaller $\beta'$,
the PBH contribution gradually decreases, approaching the results without
PBHs shown for comparison in the figures below.

\subsection{Evolution of the Boltzmann system}

\begin{figure}[t]
 \centering
\includegraphics[width=\textwidth]{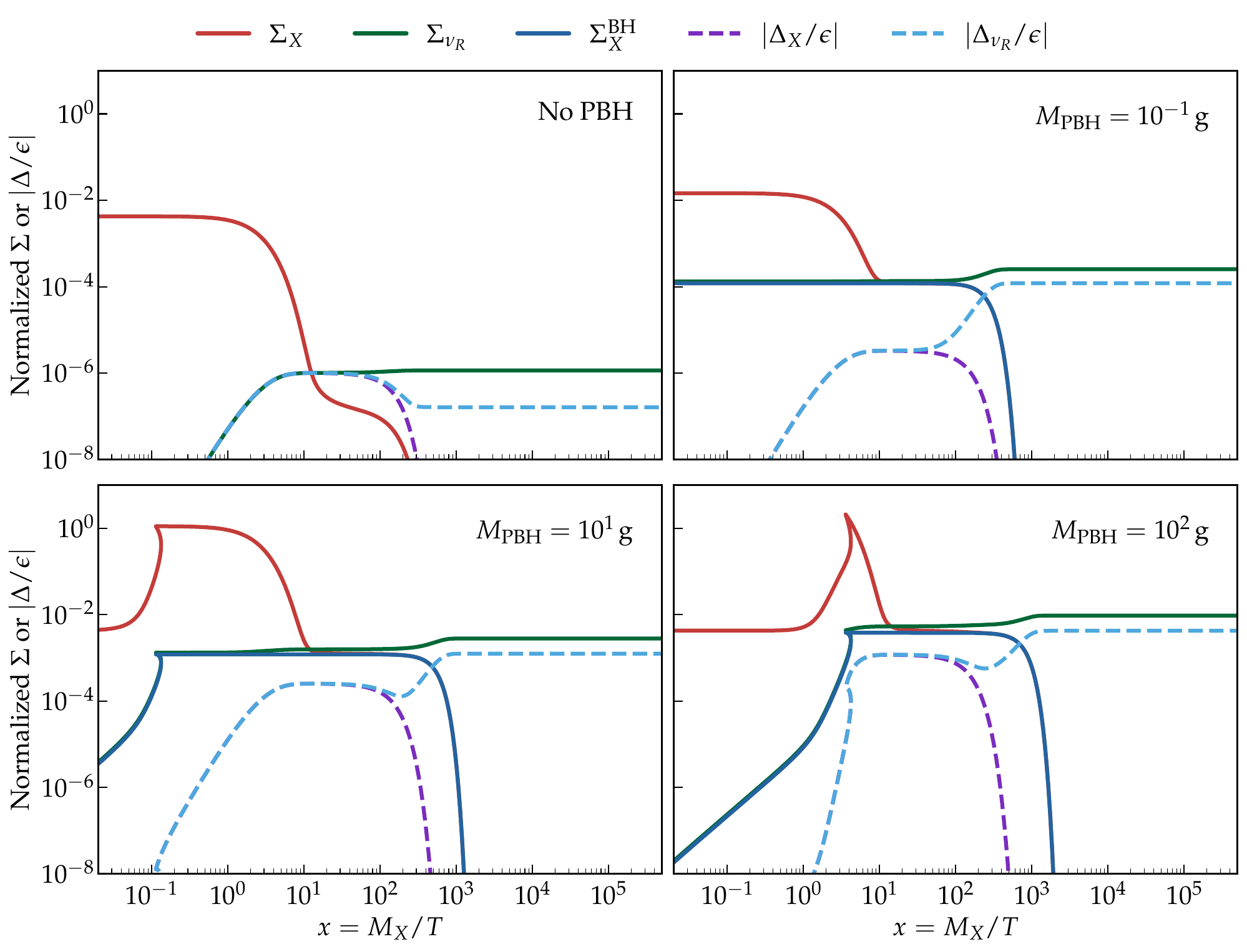}
 \caption{Evolution of the total mediator abundance
 $\Sigma_X=\Sigma_X^{\Therm}+\Sigma_X^{\BH}$, the right handed neutrino
 abundance $\Sigma_{\nu_R}$, the PBH-produced mediator component
 $\Sigma_X^{\BH}$, and the asymmetries divided by $|\epsilon|$.  All plotted
 quantities are divided by $s(T_i)/n_{\rm ref}$, corresponding to
 normalization to the initial comoving entropy.  The common
 parameters are $M_X=10^8\,{\rm GeV}$,
 $\Gamma_X/H(M_X)=10^{-4}$, $B_R=0.99$, $B_L=0.01$ and
 $\epsilon=10^{-4}$.  The PBH panels use the initial masses
 $M_{\rm PBH}=10^{-1},10^{1},10^{2}\,{\rm g}$ and $\beta'=10^{-4}$. The delayed
 rise of $\Sigma_X^{\BH}$ identifies the additional nonthermal population,
 whose CP-violating decays source the PBH contribution to the asymmetry through
 $\epsilon\Gamma_X^{\BH}\Sigma_X^{\BH}$.}
 \label{fig:boltzmann_evolution}
\end{figure}

Figure~\ref{fig:boltzmann_evolution} shows the evolution of the abundances and
asymmetries entering the Boltzmann equations.  The panel without PBHs follows
the thermal system of Ref.~\cite{Heeck:2023soj} and provides the reference
evolution, while the other three panels include the Hawking sources for $X$
and $\nu_R$.  For presentation, the numerical comoving abundances are divided
by $s(T_i)/n_{\rm ref}$ and are therefore normalized to the initial comoving
entropy, $S_{\rm ini}=s(T_i)a_i^3$, with $a_i=1$.  The period of PBH domination
is moderate for $M_{\rm PBH}=10^{-1}\,{\rm g}$ and
becomes pronounced for the two heavier cases.

In the reference case, $\Sigma_X$ follows the standard thermal evolution:
annihilations and decays gradually deplete the $X$ population as $x$
increases, while decays and inverse decays populate $\Sigma_{\nu_R}$ and
generate the two asymmetries.  Once $\Sigma_X$ has been depleted,
$\Delta_X$ approaches zero, whereas $\Delta_{\nu_R}$ remains constant.  This
sequence provides the baseline against which the additional PBH terms can be
compared.

In the remaining panels, the first modification is the source
$\mathcal S_X^{\PBH}$, which builds up $\Sigma_X^{\BH}$.  The subsequent decays
of this population feed both $\Sigma_{\nu_R}$ and $\Delta_{\nu_R}$ through
the terms proportional to $B_R\Gamma_X^{\BH}\Sigma_X^{\BH}$ and
$\epsilon\Gamma_X^{\BH}\Sigma_X^{\BH}$ in
Eqs.~\eqref{eq:pbh_even_system} and~\eqref{eq:pbh_odd_system}.  At the same
time, right handed neutrinos emitted directly by the PBHs increase
$\Sigma_{\nu_R}$ without directly sourcing $\Delta_{\nu_R}$, and also
contribute to $\Delta N_{\rm eff}$.

The timing of these contributions depends on the initial PBH mass.  For
$M_{\rm PBH}=10^{-1}\,{\rm g}$, the Hawking source acts early enough that
$\Sigma_X^{\BH}$ remains present after the thermal part of $\Sigma_X$ has
been depleted.  Its decay then produces the second increase of
$|\Delta_{\nu_R}/\epsilon|$ at $x\sim10^2$--$10^3$.  Increasing the initial
mass to $M_{\rm PBH}=10^1$ or $10^2\,{\rm g}$ lengthens the PBH lifetime and shifts
the same sequence to larger $x$, where it overlaps less with the thermal
evolution of $X$.  Across the three PBH panels, production, accumulation and
decay occur progressively later as $M_{\rm PBH}$ increases.  This shift follows
directly from $\tau_{\rm PBH}\propto M_{\rm PBH}^3$.  The Hawking contribution
is therefore a delayed source whose time profile depends on $M_{\rm PBH}$.

The delayed particle production seen for the heavier PBHs is accompanied by a
period of PBH domination before evaporation.  The corresponding evolution of
the energy fractions, including the transfer of PBH energy to radiation,
the PBH-produced $X$ population and $\nu_R$, is shown in
Appendix~\ref{app:background_evolution}.

\subsection{PBH abundance and dark radiation}

\begin{figure}[t]
 \centering
\includegraphics[width=0.72\textwidth]{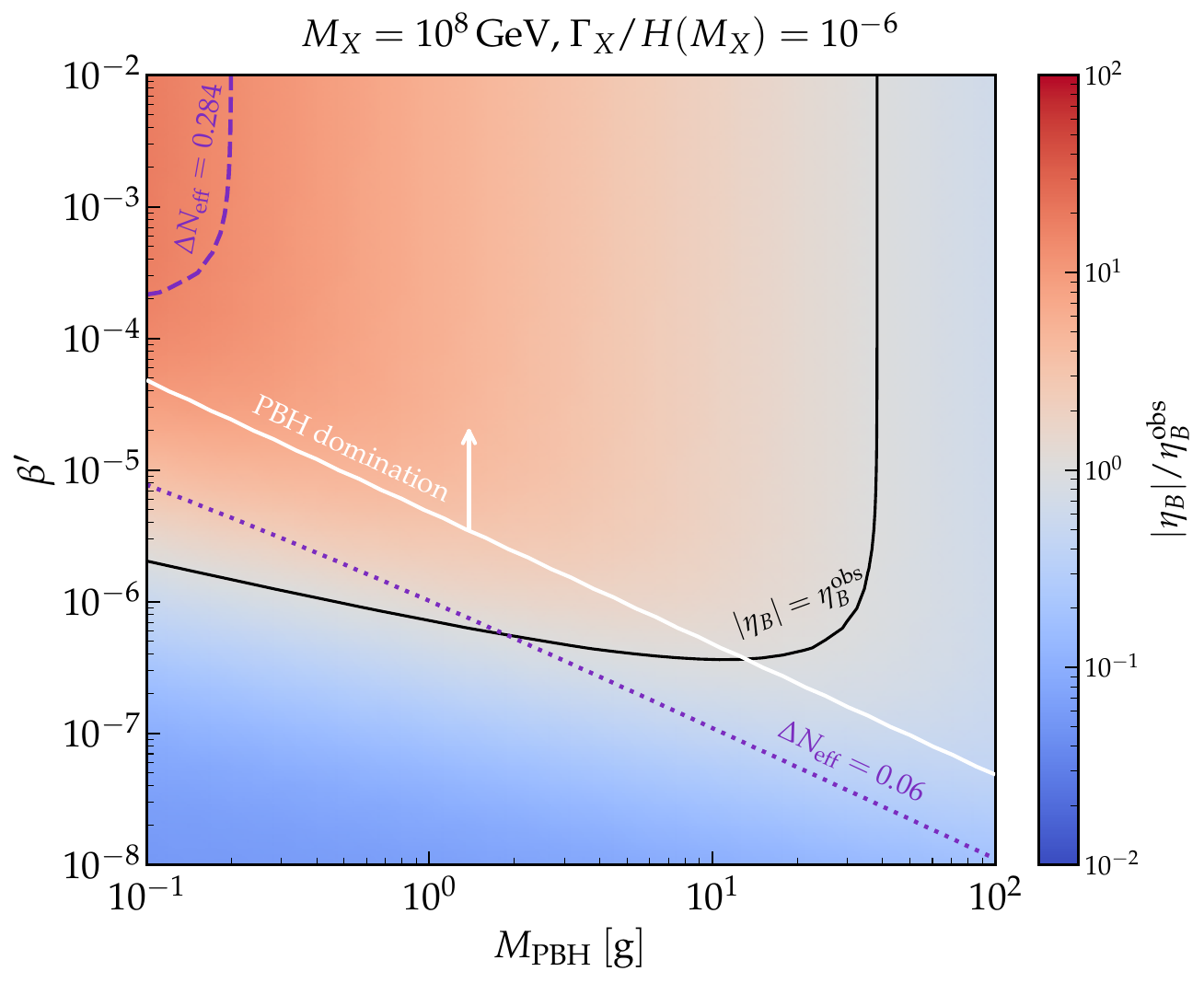}
 \caption{Scan in the initial PBH mass and abundance plane
 $(M_{\rm PBH},\beta')$ for
 $M_X=10^8\,{\rm GeV}$, $\Gamma_X/H(M_X)=10^{-6}$,
 $B_R=0.99$, $B_L=0.01$ and $\epsilon=10^{-4}$.  The color scale gives
 $|\eta_B|/\eta_B^{\rm obs}$ and the black line marks
 $|\eta_B|=\eta_B^{\rm obs}$.  The purple dashed and dotted curves show the
 Planck-era reference level $\Delta N_{\rm eff}=0.284$~\cite{Planck:2018vyg}
 and the projected CMB-S4 sensitivity $\Delta N_{\rm eff}=0.06$
 at $95\%$ confidence~\cite{Abazajian:2019eic}, respectively.  The white dashed line marks the
 numerical onset of PBH domination, defined as the initial abundance for which
 $\Omega_{\rm PBH}$ reaches $1/2$ at some point during the evolution.  Above
 this line, PBHs dominate the energy density for a period before they
 evaporate.}
 \label{fig:pbh_plane}
\end{figure}

Having identified the new source terms in the Boltzmann evolution, we next
consider their effect on the final baryon asymmetry.  We use the observed
baryon to photon ratio
$\eta_B^{\rm obs}=6.12\times10^{-10}$~\cite{Planck:2018vyg} and call a point
sufficient at the reference $|\epsilon|=10^{-4}$ when
$|\eta_B|\geq\eta_B^{\rm obs}$.  The dependence on the initial PBH parameters
is shown in Fig.~\ref{fig:pbh_plane}.  We choose $M_X=10^8\,{\rm GeV}$ and
$\Gamma_X/H(M_X)=10^{-6}$ as an example for which the thermal contribution is
insufficient, while PBH evaporation can produce the observed asymmetry.  Without PBHs,
$|\eta_B|/\eta_B^{\rm obs}\simeq0.051$, and the figure shows how this value
changes as $M_{\rm PBH}$ and $\beta'$ are varied.  Here $\beta'$
follows Eq.~\eqref{eq:beta_prime_def}, with the underlying initial fraction
defined as $\beta=\rho_{\rm PBH}/\rho_{\rm tot}$ rather than
$\rho_{\rm PBH}/\rho_R$. The black line marks
$|\eta_B|=\eta_B^{\rm obs}$.  The warmer colors identify the region in which
the PBH contribution enhances the baryon asymmetry.  For this benchmark,
gram and subgram PBHs can raise it to the observed value, whereas the
enhancement is no longer sufficient above a mass of a few tens of grams.  The
white dashed line marks the numerical onset of PBH domination.  Above this
line, $\Omega_{\rm PBH}$ exceeds $1/2$ during the evolution.
The line follows approximately the inverse-mass scaling of
Eq.~\eqref{eq:beta_prime_domination_estimate}, while its normalization is
determined from the numerical background evolution. Whether PBHs enhance the
final baryon asymmetry depends not only on whether they dominate, but also on
the timing of Hawking production, $X$ decays and entropy release.

The same scan contains the reference contours $\Delta N_{\rm eff}=0.284$ and $0.06$, calculated from Eq.~\eqref{eq:DNeff_definition}.  We retain them for comparison with the Planck-era constraint and the Stage-4 design sensitivity used in earlier PBH and Dirac-leptogenesis studies~\cite{Planck:2018vyg,Abazajian:2019eic} \footnote{A recent analysis combining primordial-abundance, CMB, and BAO measurements reports $\Delta N_{\rm eff}<0.107$ at $95\%$ confidence~\cite{Goldstein:2026iuu}; the precise constraint remains data-set and likelihood dependent. We do not include these here. }. Figure~\ref{fig:pbh_plane} therefore illustrates the correlation between baryogenesis and dark radiation rather than defining a current allowed region.
For the same benchmark, we verified that exchanging the branching ratios to
$B_R=0.01$ and $B_L=0.99$ leaves the contour
$|\eta_B|=\eta_B^{\rm obs}$ almost unchanged.  The dark radiation contribution
is instead reduced because fewer $X$ decays populate the right handed neutrino
sector.  The level
$\Delta N_{\rm eff}=0.284$ is then not reached in the displayed plane, while
$\Delta N_{\rm eff}=0.06$ requires slightly larger $\beta'$.

As discussed in Sec.~\ref{sec:dirac_review}, we use $\epsilon$ as an effective
parameter because the flavor structure of the heavier states entering the
loop is not specified.  For $B_R=0.99$ and $B_L=0.01$, our reference value
$\epsilon=10^{-4}$ lies below the bound
$|\epsilon|\leq\min(B_L,B_R)=10^{-2}$ in
Eq.~\eqref{eq:branching_epsilon_review}.  Within our Boltzmann system, and
keeping all other parameters fixed, $\epsilon$ enters only as an overall
factor in the asymmetry sources.  The final baryon asymmetry therefore scales as
$\eta_B(\epsilon)\simeq(\epsilon/10^{-4})\eta_B(10^{-4})$.  If
${\cal R}=|\eta_B|/\eta_B^{\rm obs}$ is evaluated at $\epsilon=10^{-4}$, the
observed magnitude can be obtained with $|\epsilon|\simeq10^{-4}/{\cal R}$,
provided that ${\cal R}>0$ and the required value satisfies the same bound.
For example, ${\cal R}=10,10^2$ and $10^3$ correspond to
$|\epsilon|\simeq10^{-5},10^{-6}$ and $10^{-7}$, respectively.  Its sign must
be chosen to reproduce the observed sign of the baryon asymmetry.  Large
values of ${\cal R}$ can therefore compensate for a smaller CP asymmetry,
although these sufficient asymmetry regions should still be interpreted as benchmark
regions rather than absolute predictions of a particular flavor model.  Most
of the regions shown here would no longer produce a sufficient asymmetry for
$|\epsilon|\sim10^{-7}$--$10^{-6}$.  This rescaling does not reduce the dark
radiation signal, which is CP even and essentially independent of $\epsilon$.

\subsection{Extension of the particle parameter space}

\begin{figure}[t]
 \centering
\includegraphics[width=\textwidth]{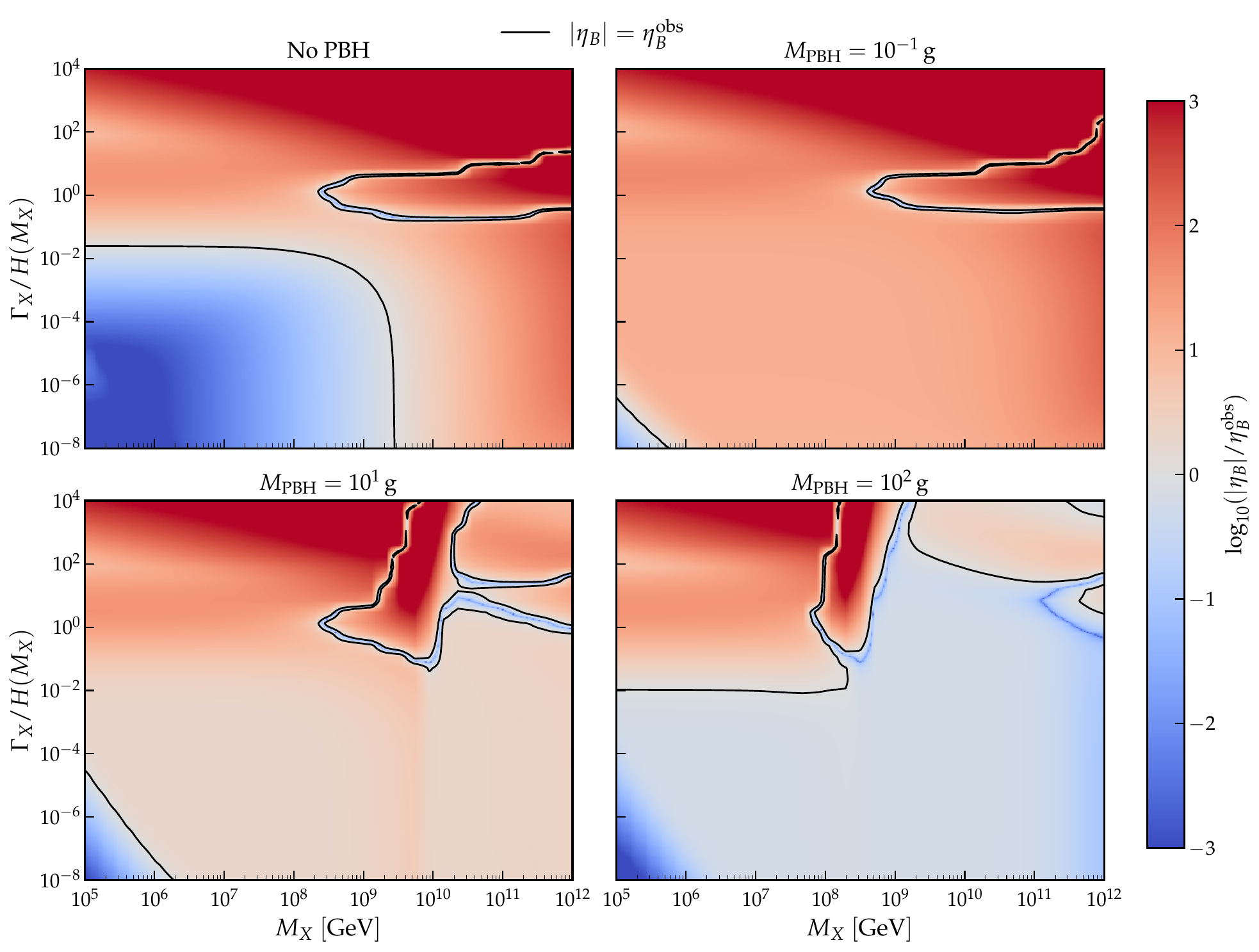}
 \caption{Final baryon asymmetry in the
 $(M_X,\Gamma_X/H(M_X))$ plane for $B_R=0.99$, $B_L=0.01$,
 $\epsilon=10^{-4}$ and $\beta'=10^{-4}$.  The color scale gives
 $\log_{10}(|\eta_B|/\eta_B^{\rm obs})$, and the black contours mark
 $|\eta_B|=\eta_B^{\rm obs}$.  The upper-left panel shows the thermal result
 without PBHs, while the other panels use
 $M_{\rm PBH}=10^{-1},10^{1},10^{2}\,{\rm g}$.  The lighter PBHs extend the
 region with sufficient asymmetry toward smaller $M_X$ at low
 $\Gamma_X/H(M_X)$, whereas the extension is smaller and more strongly
 reshaped for $M_{\rm PBH}=10^2\,{\rm g}$.  The narrow regions with small
 $|\eta_B|$ correspond to sign changes of $\eta_B$, shown explicitly in
 Appendix~\ref{app:signed_eta}.}
 \label{fig:particle_plane_etaB}
\end{figure}

Figure~\ref{fig:particle_plane_etaB} gives the main result in the particle
parameter plane.  For the PBH cases we fix $\beta'=10^{-4}$ and show
$\log_{10}(|\eta_B|/\eta_B^{\rm obs})$ for three initial PBH masses.  The
black contours mark $|\eta_B|=\eta_B^{\rm obs}$.

The panel without PBHs reproduces the structure of the thermal calculation in
Ref.~\cite{Heeck:2023soj} over the range displayed here, while the PBH panels should be compared with the general nonthermal enhancement observed in PBH-assisted leptogenesis studies~\cite{Bernal:2022pue,Ghoshal:2023fno}.  For
$\Gamma_X/H(M_X)\lesssim10^{-2}$, the thermal calculation without PBHs
produces a sufficient asymmetry only for $M_X$ around $10^9\,{\rm GeV}$ or
above.  PBHs with initial masses $M_{\rm PBH}=10^{-1}\,{\rm g}$ and
$10^1\,{\rm g}$ extend this region down to
$M_X\sim10^6\,{\rm GeV}$. Within the approximations adopted here,
these benchmarks show that PBHs can lower the required scale by up to roughly
three orders of magnitude.  The
extension is smaller for $M_{\rm PBH}=10^2\,{\rm g}$, since the later Hawking
emission is less effective in the low-mass part of the scan.
This behavior is consistent with Eq.~\eqref{eq:pbh_threshold_mass}, which
estimates when Hawking emission of $X$ becomes efficient. Crossing this
threshold is not sufficient by itself: the contribution to the final
asymmetry also depends on the decay timing of the emitted $X$ population and
on the entropy released during evaporation. This explains why
$M_{\rm PBH}=10^2\,{\rm g}$ produces a smaller extension than the lighter PBH
cases.

The opposite behavior appears where the thermal calculation already produces
a large asymmetry.  In these regions the PBH contribution does not simply add
to the thermal result: Hawking production, its different decay timing and
entropy injection can reshape or reduce the final $|\eta_B|$.  For the
reference value $\epsilon=10^{-4}$, some of these points still remain above
the observed value despite this reduction.

Appendix~\ref{app:branching_ratio} repeats this scan for the complementary
choice $B_R=0.01$ and $B_L=0.99$.  The PBH-assisted extension at
$\Gamma_X/H(M_X)\lesssim10^{-2}$ remains visible, showing that the reduction
of the required $X$ mass is not specific to $B_R=0.99$.  The contours at
larger rates are instead reshaped because the branching fractions modify the
transfer of particles and asymmetries between the left- and right handed
sectors.

The narrow valleys and branches in the PBH panels are not failed numerical
points.  They occur where the signed asymmetry crosses zero because the
thermal source, the PBH-produced $X$ source and washout dominate at different
stages of the evolution.  The corresponding signed maps are shown in
Appendix~\ref{app:signed_eta}, where these zero crossings can be followed
explicitly.  Since the final asymmetry is proportional to $\epsilon$, changing
the sign of $\epsilon$ reverses the sign of $\eta_B$.  We therefore show
$|\eta_B|$ in the main scan.
We do not add the separate sphaleron-decoupling region shown in Fig.~2 of
Ref.~\cite{Heeck:2023soj}.  In our calculation the baryon asymmetry is fixed at
$T_{\rm sph}$ through Eq.~\eqref{eq:etaB_observable}, while the later evolution
accounts only for entropy dilution.

\begin{figure}[t]
 \centering
\includegraphics[width=\textwidth]{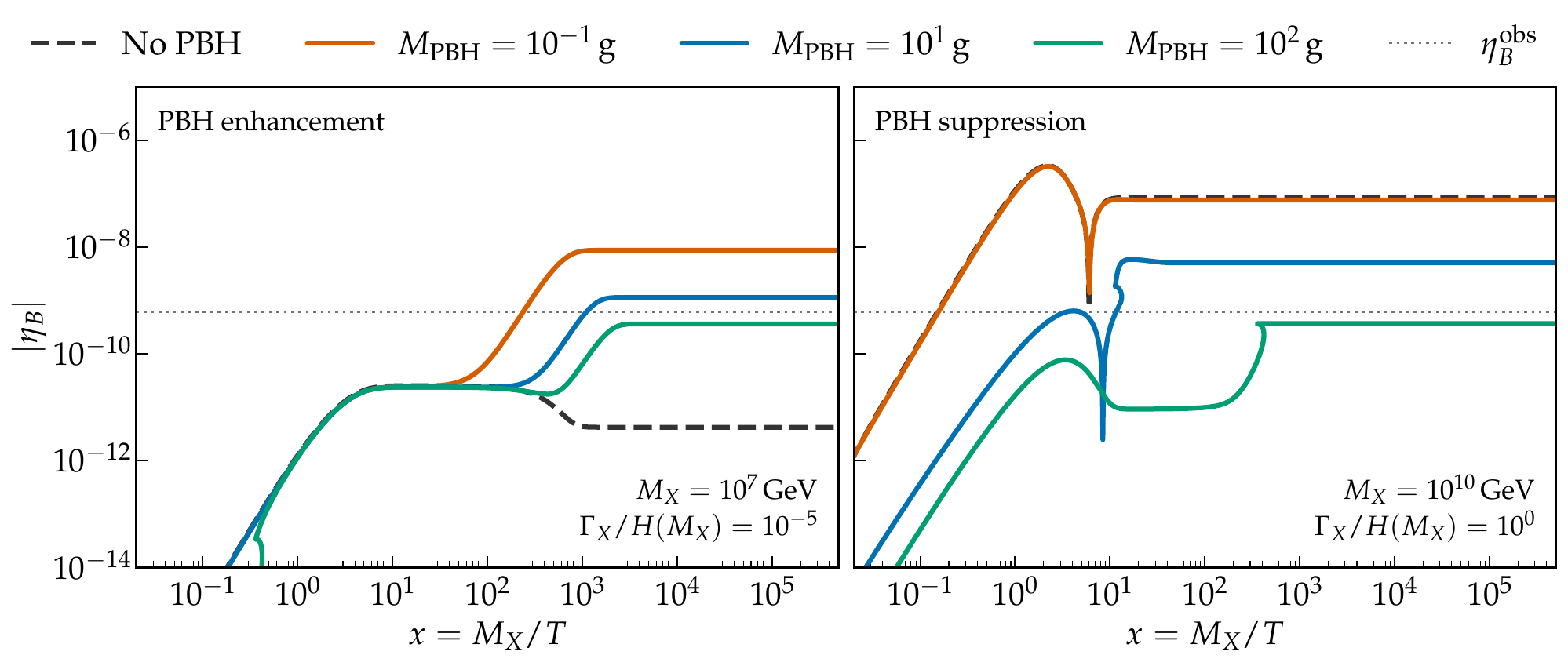}
 \caption{Evolution of $|\eta_B|$ in two regions of the particle parameter
 plane for $B_R=0.99$, $B_L=0.01$, $\epsilon=10^{-4}$ and
 $\beta'=10^{-4}$.  The left panel uses $M_X=10^7\,{\rm GeV}$ and
 $\Gamma_X/H(M_X)=10^{-5}$, where PBHs enhance an insufficient thermal
 asymmetry.  The $10^{-1}\,{\rm g}$ and $10^1\,{\rm g}$ cases exceed the
 observed value, while the $10^2\,{\rm g}$ case remains below it.  The right
 panel uses $M_X=10^{10}\,{\rm GeV}$ and $\Gamma_X/H(M_X)=1$, where the
 thermal asymmetry is already sufficient and the PBH cases leave it unchanged
 or reduce it.  For the parameters shown, the $10^{-1}\,{\rm g}$ and
 $10^1\,{\rm g}$ cases remain above the observed value, whereas the
 $10^2\,{\rm g}$ case falls below it.  The sharp dips mark zero crossings of
 the signed asymmetry, which appear as minima because $|\eta_B|$ is plotted.
 The dotted line denotes $\eta_B^{\rm obs}$.}
 \label{fig:etaB_evolution_benchmarks}
\end{figure}

Figure~\ref{fig:etaB_evolution_benchmarks} follows the evolution at two
representative locations chosen to illustrate the two possible effects.  In
the left panel the thermal result is insufficient and all three PBH cases
increase the final asymmetry.  The $M_{\rm PBH}=10^{-1}\,{\rm g}$ and
$10^1\,{\rm g}$ curves rise above the observed value, while the
$10^2\,{\rm g}$ case remains below it.  In the right panel the thermal result
is already above the observed value.  The lightest PBH case stays close to
the thermal curve, while the two larger masses reduce the final asymmetry.
For $\epsilon=10^{-4}$, the $10^{-1}\,{\rm g}$ and $10^1\,{\rm g}$ PBH cases
remain above $\eta_B^{\rm obs}$, whereas the $10^2\,{\rm g}$ case falls below
it.  The position of individual curves relative to the observed value depends
on the benchmark choice, in particular on $\epsilon$; the relevant point is
that PBHs can enhance the asymmetry in one region and reduce it in another.
The sharp dips occur when $\eta_B$ crosses zero and reappears with the
opposite sign.

To identify the origin of the PBH effect, we repeated the two benchmark
calculations by first including only the modified expansion and entropy
evolution, without Hawking production of $X$ or direct emission of $\nu_R$.
For $M_{\rm PBH}=10^{-1}\,{\rm g}$, this gives almost the same final asymmetry as
the calculation without PBHs.  The enhancement appears when Hawking
production of $X$ is included and accounts for nearly all of the change.
Directly emitted $\nu_R$ modify the final asymmetry by about $3\%$ in the left
benchmark and $0.5\%$ in the right one.  Their effect on $\eta_B$ is therefore
small for these two points, but their energy density contributes directly to
$\Delta N_{\rm eff}$ and remains relevant for the dark radiation signal.
This illustrates why $\eta_B$ and $\Delta N_{\rm eff}$ provide complementary
information.  The former depends on the signed asymmetry $\Delta_{\nu_R}$,
whereas the latter depends on the total energy stored in
$\nu_R+\bar\nu_R$.  Direct symmetric emission can therefore affect dark
radiation even when its effect on $\eta_B$ is small.

We restrict the scan to $M_X\leq10^{12}\,{\rm GeV}$ and
$\Gamma_X/H(M_X)\leq10^4$ in order to focus on the region where the PBH
contribution is most relevant.  Without PBHs, the region with sufficient
asymmetry extends beyond $M_X=10^{12}\,{\rm GeV}$, while Hawking production
becomes less effective for heavier
$X$ and larger $M_{\rm PBH}$, so the right edge is not a physical boundary.
Similarly, we stop at $\Gamma_X/H(M_X)=10^4$ because the scattering
contribution discussed in Sec.~\ref{sec:dirac_review} grows with $\Gamma_X$
and becomes noticeable near the upper-right edge of the scan.  The displayed
plane should therefore be read as the domain relevant for the PBH effect and
for the decay and inverse-decay approximation used here
\footnote{Due to the uncertainties involved, possible interactions involving the PBH-produced mediator population could
modify the size of the enhancement and slightly shift the numerical boundaries of the displayed regions.}.  The PBH effect is
largest where the thermal mechanism alone produces too little asymmetry.
There, Hawking emission opens viable parameter space.  Where the thermal
mechanism is already efficient, PBHs mainly reshape or reduce the existing
asymmetry.

\section{Discussion}
\label{sec:discussion}

We now place the numerical results in a broader context.  We first compare the
mechanism with PBH-assisted Majorana leptogenesis, focusing on the physical
differences without attempting a direct parameter mapping.  We then
discuss the possible connection between the ultralight PBH population
considered here and primordial gravitational-wave signals.

\subsection{Comparison with PBH-assisted Majorana leptogenesis}
\label{sec:majorana_comparison}

In both the Dirac and Majorana scenarios, Hawking evaporation can populate a
heavy unstable state even when its thermal production is inefficient.  The
subsequent charge dynamics are different, and a direct comparison of the two
parameter spaces is not possible, since the emitted
particles and the microscopic parameters controlling the asymmetry are
different.

In both cases, the parent state becomes unsuppressed in the Hawking spectrum
when $T_{\rm BH}$ reaches its mass.  Equation~\eqref{eq:pbh_threshold_mass}
gives this threshold for $X$; the corresponding Majorana estimate follows by
replacing $M_X$ with the heavy-neutrino mass.  In PBH-assisted Majorana
leptogenesis based on heavy right handed neutrinos, the emitted particle is
itself the out-of-equilibrium parent whose lepton-number-violating decay
generates the asymmetry.  Production at a low plasma temperature can avoid
the strong washout that would occur in the standard thermal history
\cite{Perez-Gonzalez:2020vnz,Bernal:2022pue,Barman:2022gjo,
Calabrese:2023key,Schmitz:2023pfy,Calabrese:2023bxz,
Barman:2024slw,Calabrese:2025sfh,Ghoshal:2023fno}.  In the Dirac mechanism
studied here, PBHs instead produce the charged scalar $X$.  Its decays
conserve total lepton number but generate opposite asymmetries in the visible
and right handed neutrino sectors.

The CP asymmetries also have different microscopic origins.  In the
heavy-Majorana-neutrino scenarios used for comparison, the CP asymmetry
depends on the neutrino couplings and mass spectrum, whereas in the Dirac
setup considered here $\epsilon$ arises from the heavier charged scalars
entering the loop contribution in
Eq.~\eqref{eq:epsilon_parametric_revision}.

In both cases, the final baryon asymmetry depends on the abundance produced by
the PBHs, the relevant CP asymmetry, washout and the entropy released during
evaporation.  The role of Hawking production is different, however.  In the
Majorana case it can produce the asymmetry-generating right handed neutrinos
after thermal washout has become inefficient.  In the Dirac case it supplies
an additional population of $X$ where the thermal mechanism alone produces
too little asymmetry.

The reduction of the required $X$ mass found in
Fig.~\ref{fig:particle_plane_etaB} is therefore the Dirac counterpart of the
opening of parameter space found in PBH-assisted Majorana leptogenesis.  This
comparison is only at the level of the mechanism and does not imply a direct
mapping between $M_N$ and $M_X$, or between their CP asymmetries.  A
distinctive feature of the Dirac scenario is that the populated $\nu_R$
sector also contributes to $\Delta N_{\rm eff}$, providing an additional
cosmological observable.

\subsection{Gravitational-wave probes of ultralight PBH-assisted leptogenesis}
\label{sec:gw_probes}

The ultralight PBHs considered in
Figs.~\ref{fig:pbh_plane}-\ref{fig:etaB_evolution_benchmarks} evaporate long
before BBN, but their formation history may leave a high-frequency primordial
GW signal~\cite{Ghoshal:2023fno,Borah:2026zbl}.  For formation during radiation domination, the horizon scale maps approximately to the present frequency $f_{\rm PBH}$,
\begin{equation}
  f_{\rm PBH}\simeq3.8\times10^{8}\,{\rm Hz}
  \left(\frac{\gamma}{0.2}\right)^{1/2}
  \left(\frac{106.75}{g_\star(T_{\rm f})}\right)^{1/12}
  \left(\frac{1\,{\rm g}}{M_{\rm PBH}}\right)^{1/2}.
  \label{eq:gw_frequency_mass_relation}
\end{equation}
Thus $M_{\rm PBH}=10^{-1}$--$10^2\,{\rm g}$ corresponds to $f_{\rm PBH}\sim10^7$--$10^9\,{\rm Hz}$.  If enhanced curvature perturbations create the PBHs, the same perturbations source second-order GWs, so their peak frequency fixes the PBH mass scale while their amplitude constrains the fluctuations that determine $\beta'$~\cite{Ananda:2006af,Baumann:2007zm,Kohri:2018awv,Inomata:2016rbd}.  This directly complements the baryogenesis scan, in which $(M_{\rm PBH},\beta')$ controls Hawking production and entropy release.

If PBHs dominate before evaporation, the intervening matter-dominated era and the return to radiation domination reshape the GW transfer function~\cite{Inomata:2020lmk}.  Defining the comoving SM entropy growth factor as $\Delta_{\rm ent}\equiv S_{\rm after}/S_{\rm before}$, entropy injection acts on pre-existing GWs and on the baryon asymmetry according to the useful scaling
\begin{equation}
  \Omega_{\rm GW}^{\rm pre}\longrightarrow
  \Omega_{\rm GW}^{\rm pre}\Delta_{\rm ent}^{-4/3},
  \qquad
  \eta_B\longrightarrow\eta_B\Delta_{\rm ent}^{-1}.
  \label{eq:gw_entropy_dilution_scaling}
\end{equation}
Such a gravitational-wave signal would constrain the PBH formation and
expansion history.  Its connection to the Dirac scenario considered here
would have to be assessed together with the baryon asymmetry and the correlated
$\Delta N_{\rm eff}$ contribution from the populated right handed neutrino
sector.  Additional estimates for induced GWs, evaporation-scale features, defect backgrounds, blue-tilted tensors and the comparison with Majorana leptogenesis are collected in Appendix~\ref{app:gw_details}.

\section{Conclusions} \label{sec:conclusion}
We have analyzed, for the first time to our knowledge, the impact of an evaporating PBH population on the charged-mediator realization of Dirac leptogenesis.  We have solved the coupled Boltzmann equations for the thermal
$X$ population, the $X$ particles emitted by PBHs, the $\nu_R$ sector and the
cosmological background, including Hawking production, washout, the modified
expansion rate and the entropy released during evaporation.  This allows us
to follow the timing of the different sources and their decays relative to
the thermal evolution and entropy release.

As shown in Fig.~\ref{fig:boltzmann_evolution}, PBH evaporation supplies an
additional population of $X$ after thermal production has become inefficient.
Figure~\ref{fig:pbh_plane} places this effect in the initial PBH mass and
abundance plane.  The outcome depends on both the number of PBHs and their
evaporation time, since Hawking production, the expansion history and entropy
injection all contribute to the final asymmetry.

The main result is displayed in Figs.~\ref{fig:particle_plane_etaB}
and~\ref{fig:etaB_evolution_benchmarks}.  In the parameter range explored,
PBHs have their largest impact where the thermal source is inefficient,
lowering the required $X$ mass by up to three orders of magnitude and allowing
a smaller effective CP asymmetry through the approximate scaling with
$\epsilon$.  Where the thermal source is already efficient, however, later
decays and entropy release can reduce the asymmetry.  The interplay between
the thermal and PBH contributions also produces the sign changes and
structures visible in the particle parameter plane.

The right handed neutrino sector provides a complementary connection to dark
radiation.  The baryon asymmetry depends on the signed $\nu_R$ asymmetry at
sphaleron freeze-out, whereas $\Delta N_{\rm eff}$ measures the total energy
stored in $\nu_R$ and $\bar\nu_R$.  Direct Hawking emission of right handed
neutrinos therefore contributes to the latter without generating a signed
asymmetry.  As Fig.~\ref{fig:pbh_plane} shows, part of the region with
sufficient asymmetry predicts
$0.06<\Delta N_{\rm eff}<0.284$, between the projected CMB-S4 sensitivity and
the displayed Planck-era reference,
offering a possible probe of the associated $\nu_R$ population even though
such a signal would not identify its PBH origin on its own.

Our treatment follows the PBH-produced mediator population through
its number and energy densities without resolving its full momentum
distribution.  Its interactions with the plasma, extended PBH mass
distributions, rotating PBHs and explicit flavor models for the effective CP
asymmetry remain for future work and may shift the precise
boundaries.  Within the present treatment, Hawking
production of $X$ supplies an efficient source where the thermal mechanism
alone produces too little asymmetry, while the energy deposited in $\nu_R$
leaves a separate contribution
to $\Delta N_{\rm eff}$.

\clearpage

\appendix
\section{Cosmological background evolution}
\label{app:background_evolution}

\begin{figure}[t]
 \centering
\includegraphics[width=\textwidth]{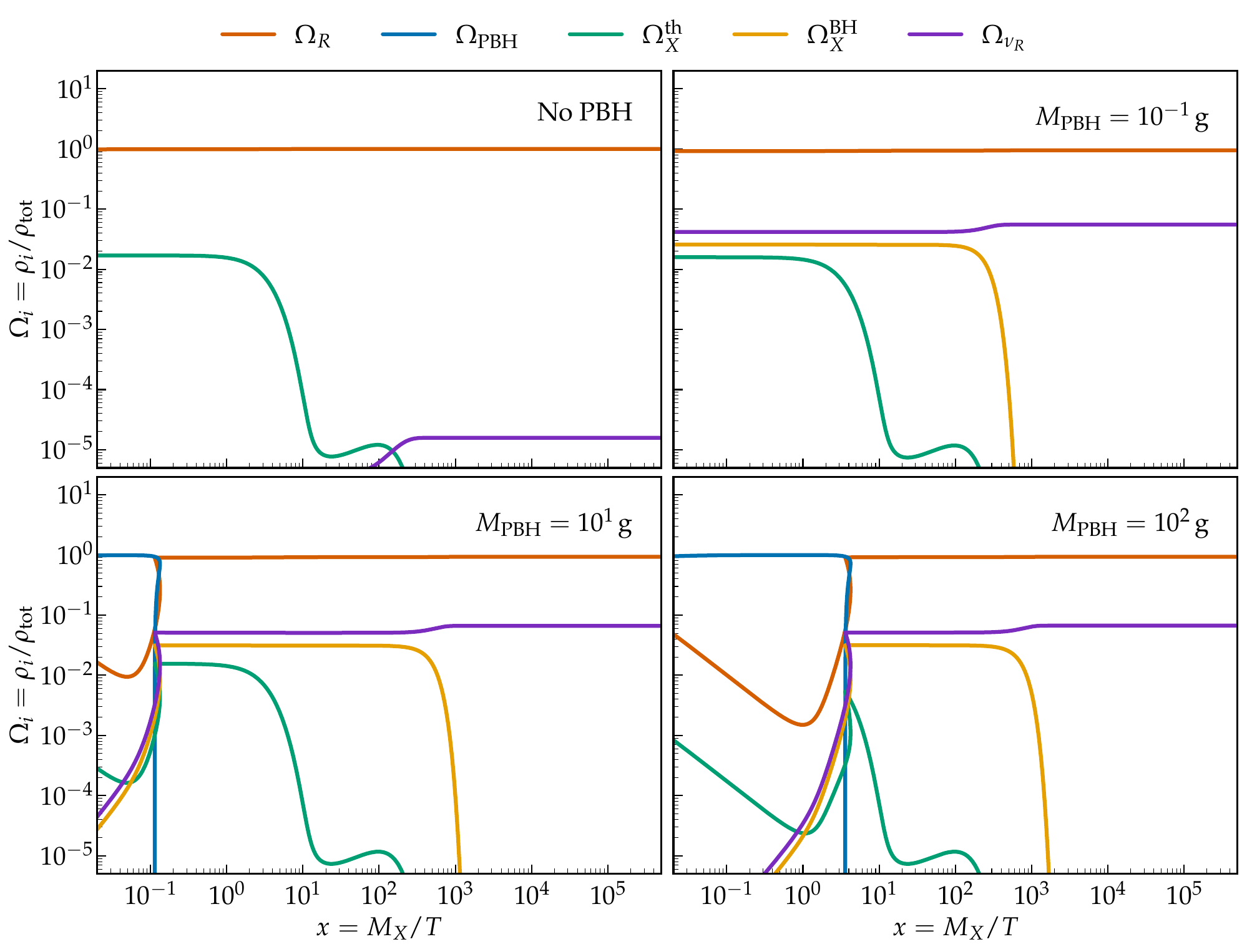}
 \caption{Evolution of the energy density fractions
 $\Omega_j\equiv\rho_j/\rho_{\rm tot}$ for the SM radiation bath, PBHs,
 the thermally produced and PBH-produced $X$ populations, and right handed
 neutrinos.  The common parameters are
 $M_X=10^8\,{\rm GeV}$, $\Gamma_X/H(M_X)=10^{-4}$, $B_R=0.99$, $B_L=0.01$,
 $\epsilon=10^{-4}$ and $\beta'=10^{-4}$.  The panels show the case without
 PBHs and those with initial masses
 $M_{\rm PBH}=10^{-1},10^{1},10^{2}\,{\rm g}$.  The $10^1$ and $10^2\,{\rm g}$
 cases are initially PBH dominated in the displayed interval, whereas the
 $10^{-1}\,{\rm g}$ PBHs have already evaporated.}
 \label{fig:energy_fractions_appendix}
\end{figure}

Figure~\ref{fig:energy_fractions_appendix} shows the energy budget for the
same benchmark as Fig.~\ref{fig:boltzmann_evolution}.  Without PBHs, radiation
remains the dominant component while the thermal $X$ population decays and a
small fraction of the energy is transferred to $\nu_R$.  For
$M_{\rm PBH}=10^1$ and $10^2\,{\rm g}$, the evolution enters the displayed range
with $\Omega_{\PBH}$ close to unity.  Evaporation then restores radiation
domination and feeds the PBH-produced $X$ and $\nu_R$ populations.  The
$10^2\,{\rm g}$ case makes this sequence particularly clear because the
evaporation occurs at larger $x$.  For $M_{\rm PBH}=10^{-1}\,{\rm g}$, the PBHs
have already evaporated before the interval shown.

This evolution also gives a direct interpretation of the white line in
Fig.~\ref{fig:pbh_plane}.  Above that line, $\Omega_{\PBH}$ reaches $1/2$ at
some point before evaporation.  The energy fractions therefore provide the
dynamical counterpart of the numerical PBH domination criterion used in the
PBH parameter scan.

\clearpage

\section{Sign of the baryon asymmetry}
\label{app:signed_eta}

\begin{figure}[t]
 \centering
\includegraphics[width=\textwidth]{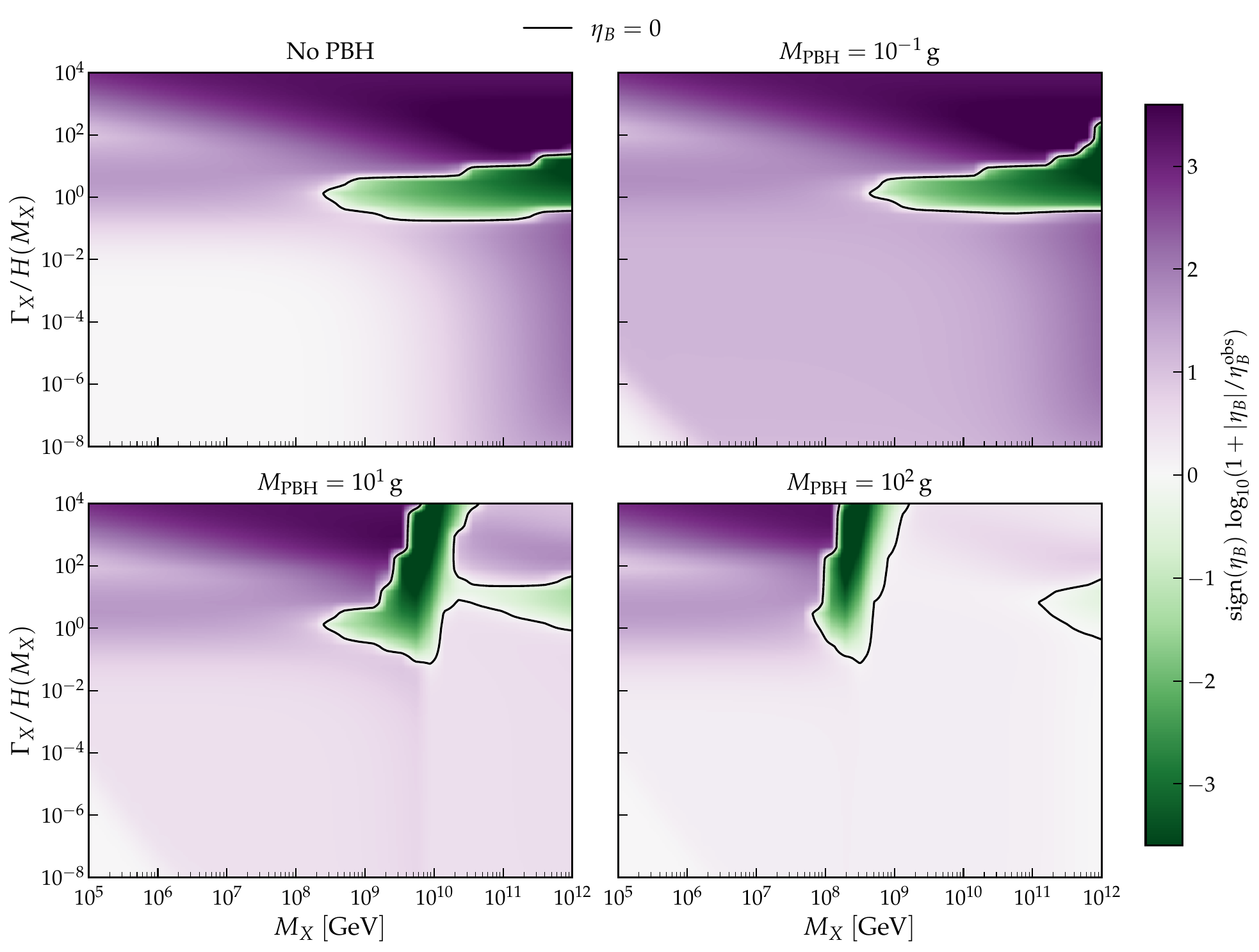}
 \caption{Signed baryon asymmetry in the
 $(M_X,\Gamma_X/H(M_X))$ plane for $B_R=0.99$, $B_L=0.01$,
 $\epsilon=10^{-4}$ and $\beta'=10^{-4}$.  The color scale shows
 ${\rm sign}(\eta_B)\log_{10}(1+|\eta_B|/\eta_B^{\rm obs})$: purple regions
 have positive $\eta_B$ and green regions negative $\eta_B$.  The solid lines
 mark $\eta_B=0$.  The first panel has no PBHs, while the other panels use
 $M_{\rm PBH}=10^{-1},10^{1},10^{2}\,{\rm g}$.  These zero contours account for
 the regions of small $|\eta_B|$ in Fig.~\ref{fig:particle_plane_etaB}.}
 \label{fig:signed_eta_appendix}
\end{figure}

The main scan uses $|\eta_B|$ to identify regions with sufficient final
magnitude.  Figure~\ref{fig:signed_eta_appendix} instead retains the sign of
the asymmetry.  The no-PBH panel reproduces the sign structure of the thermal
calculation in Ref.~\cite{Heeck:2023soj}.  Once PBHs are included, the
additional $X$ population changes the relative timing of the source and
washout terms, shifting and bending the surfaces on which the final asymmetry
vanishes.  This effect is modest for $M_{\rm PBH}=10^{-1}\,{\rm g}$ and much more
pronounced for the two heavier masses.

The solid contours in Fig.~\ref{fig:signed_eta_appendix} coincide with the
regions of small $|\eta_B|$ in Fig.~\ref{fig:particle_plane_etaB}.  The
structures in the main scan therefore arise from physical cancellations
between contributions generated at different stages of the evolution.  They
are not isolated numerical failures.  We use $\epsilon=10^{-4}$ in the figure;
reversing its sign reverses the sign of $\eta_B$ without changing its
magnitude.
Between sampled points, the signed $\eta_B$ is linearly
interpolated in $\log_{10}M_X$ and
$\log_{10}[\Gamma_X/H(M_X)]$, and the solid curves are its zero contours.

\clearpage

\section{Complementary branching-ratio choice}
\label{app:branching_ratio}

\begin{figure}[t]
 \centering
 \includegraphics[width=\textwidth]{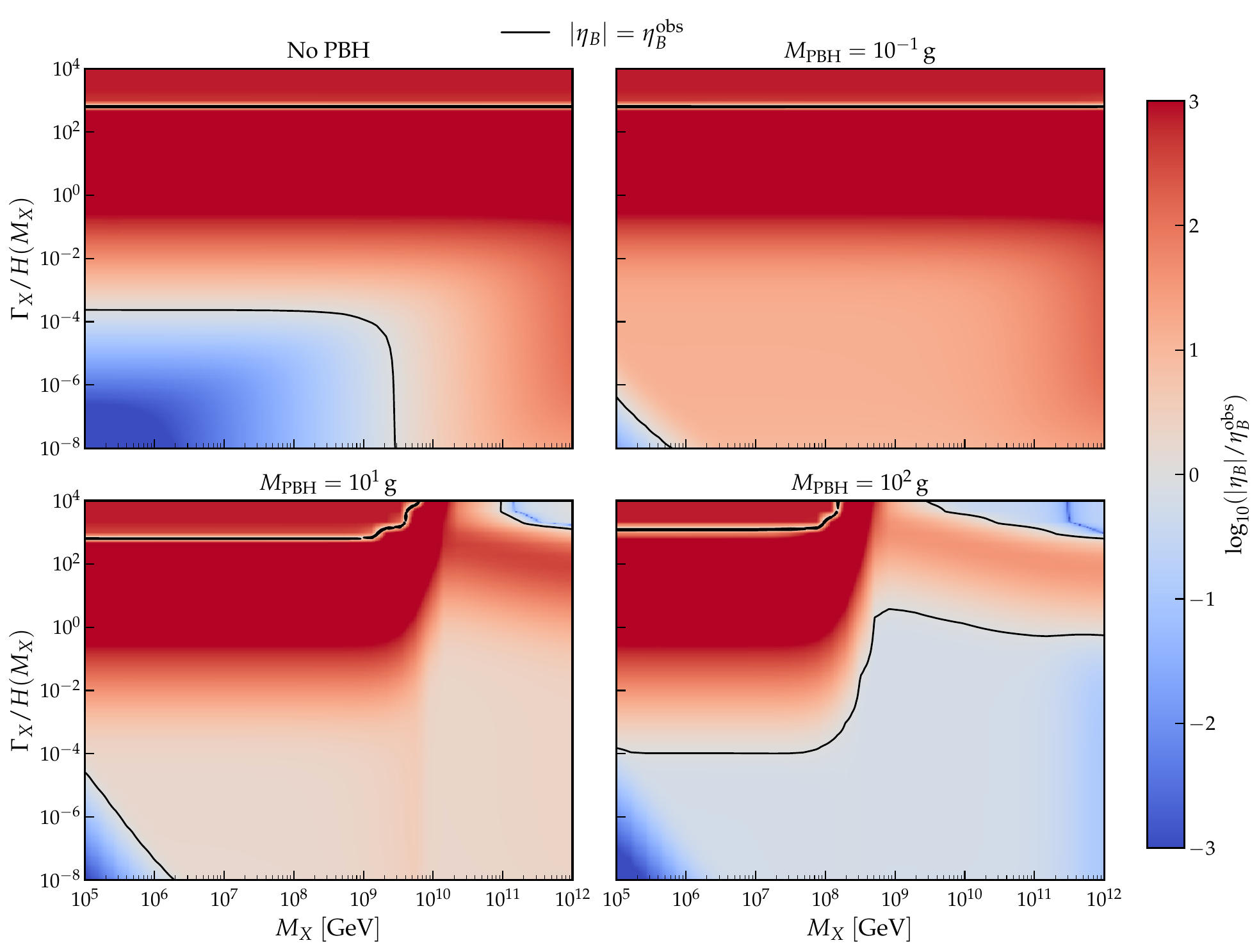}
 \caption{Final baryon asymmetry in the
 $(M_X,\Gamma_X/H(M_X))$ plane for the complementary branching-ratio choice
 $B_R=0.01$, $B_L=0.99$, with $\epsilon=10^{-4}$ and $\beta'=10^{-4}$.
 The color scale gives $\log_{10}(|\eta_B|/\eta_B^{\rm obs})$ and the black
 contours mark $|\eta_B|=\eta_B^{\rm obs}$.  The first panel has no PBHs,
 while the other panels use
 $M_{\rm PBH}=10^{-1},10^{1},10^{2}\,{\rm g}$.  As in the main scan, the
 $10^{-1}$ and $10^1\,{\rm g}$ PBHs give the clearest extension toward smaller
 $M_X$ at low $\Gamma_X/H(M_X)$.}
 \label{fig:branching_ratio_appendix}
\end{figure}

Figure~\ref{fig:branching_ratio_appendix} repeats the particle parameter scan presented in the main text, but with the complementary branching ratio choice
$B_R=0.01$, $B_L=0.99$. The qualitative PBH effect persists.  At
small $\Gamma_X/H(M_X)$, the calculation without PBHs requires a much larger
$M_X$, whereas PBHs with initial masses $10^{-1}$ and $10^1\,{\rm g}$ extend the
region with sufficient asymmetry toward substantially smaller masses.  The
$10^2\,{\rm g}$ case again has a more limited effect because its later
evaporation makes dilution comparatively more important.

The exchange of $B_R$ and $B_L$ nevertheless changes the detailed boundaries,
especially at larger decay rates and near the cancellation regions.  This is
expected because the branching fractions determine how the $X$ decays feed
the visible and right handed sectors and enter the decay and inverse-decay
terms~\cite{Heeck:2023soj}.  Despite these differences, both choices show the
PBH-assisted extension toward smaller $M_X$, while the precise contours depend
on $B_R$ and $B_L$.

\clearpage

\section{Gravitational-wave estimates}
\label{app:gw_details}

If PBHs originate from a narrow enhancement of the curvature spectrum,
$\mathcal P_\zeta(k)\simeq A_\zeta\delta(\ln k-\ln k_\star)$, the associated
second-order tensor background scales as~\cite{Ananda:2006af,Baumann:2007zm,Kohri:2018awv,Inomata:2016rbd}
\begin{equation}
  \Omega_{\rm GW,0}^{\rm ind}(f_{\rm PBH})
  \sim C_{\rm ind}\Omega_{r,0}A_\zeta^2,
  \qquad C_{\rm ind}=\mathcal O(10^{-1}\text{--}1),
  \label{eq:induced_gw_scaling}
\end{equation}
where $\Omega_{r,0}$ is the present radiation fraction and $C_{\rm ind}$
collects the shape and transfer-function dependence.  A schematic Gaussian
estimate gives
\begin{equation}
  \beta(M_{\rm PBH})\simeq\frac12\operatorname{erfc}
  \!\left(\frac{\delta_c}{\sqrt{2}\sigma_\delta}\right),
  \qquad \sigma_\delta^2\simeq\frac{16}{81}A_\zeta,
  \label{eq:beta_curvature_relation}
\end{equation}
where $\sigma_\delta^2$ is the variance of the smoothed density contrast and
$\delta_c=\mathcal O(0.4)$ is the collapse threshold, up to profile,
window-function and non-Gaussian corrections.  Thus an induced-GW amplitude
constrains the same primordial input that determines $\beta'$ in
Fig.~\ref{fig:pbh_plane}, while the peak frequency constrains the PBH mass
through Eq.~\eqref{eq:gw_frequency_mass_relation}.

The connection to the baryon asymmetry must, however, retain the full source
and washout history.  To make this explicit, define the vector of charge-odd
comoving variables
\begin{equation}
  \boldsymbol\Delta(a)
  \equiv
  \begin{pmatrix}\Delta_X(a)\\[1mm]\Delta_{\nu_R}(a)\end{pmatrix},
  \qquad
  aH\frac{d\boldsymbol\Delta}{da}
  =\boldsymbol S_{\rm CP}(a)+\boldsymbol{\mathsf W}(a)\boldsymbol\Delta(a),
  \label{eq:gw_asymmetry_vector}
\end{equation}
where $\boldsymbol{\mathsf W}$ contains the transfer and washout coefficients
in Eq.~\eqref{eq:pbh_odd_system}.  The $\nu_R$ component of the explicit CP
source is
\begin{equation}
  S_{{\rm CP},\nu_R}(a)=\epsilon\left[
  \Gamma_X^{\rm th}
  \left(\Sigma_X^{\rm th}
  -\Sigma_X^{\rm eq}\frac{\Sigma_{\nu_R}}{\Sigma_{\nu_R}^{\rm eq}}\right)
  +\Gamma_X^{\rm BH}\Sigma_X^{\rm BH}\right].
  \label{eq:gw_dirac_source}
\end{equation}
If $\boldsymbol{\mathsf U}(a,a')$ denotes the evolution operator generated by
$\boldsymbol{\mathsf W}/(aH)$, the formal solution at sphaleron freeze-out is
\begin{equation}
  \boldsymbol\Delta(a_{\rm sph})
  =\boldsymbol{\mathsf U}(a_{\rm sph},a_i)\boldsymbol\Delta(a_i)
  +\int_{a_i}^{a_{\rm sph}}\frac{da'}{a'H(a')}
  \boldsymbol{\mathsf U}(a_{\rm sph},a')\boldsymbol S_{\rm CP}(a').
  \label{eq:gw_eta_relation}
\end{equation}
Equation~\eqref{eq:gw_eta_relation}, together with
Eq.~\eqref{eq:etaB_observable}, replaces an estimate based on the mediator
abundance at a single temperature.  In particular,
$\Sigma_X^{\rm BH}(a)=a^3(n_X^{\rm BH}+n_{\bar X}^{\rm BH})/n_{\rm ref}$ is the
same fixed-reference comoving abundance used in the main calculation.  We do
not introduce a separate $Y_X^{\rm PBH}$.  If desired, an instantaneous
entropy-normalized yield is related to it by
\begin{equation}
  Y_X^{\rm BH}(a)
  \equiv\frac{n_X^{\rm BH}+n_{\bar X}^{\rm BH}}{s(T)}
  =\frac{n_{\rm ref}}{a^3s(T)}\Sigma_X^{\rm BH}(a),
  \label{eq:gw_YX_Sigma_conversion}
\end{equation}
but $Y_X^{\rm BH}$ varies during entropy injection even at fixed comoving
particle number and is not used as a proxy for $\eta_B$.

If PBHs dominate before evaporation, the intervening matter-dominated era and
the return to radiation domination reshape the GW transfer
function~\cite{Inomata:2020lmk}.  Defining the comoving SM entropy growth as
\begin{equation}
  \Delta_{\rm ent}\equiv\frac{S_{\rm after}}{S_{\rm before}},
  \qquad S\equiv s(T)a^3,
  \label{eq:gw_entropy_definition}
\end{equation}
a freely propagating GW background generated before the entropy release obeys
the schematic scaling
\begin{equation}
  \Omega_{\rm GW}^{\rm pre}\longrightarrow
  \Omega_{\rm GW}^{\rm pre}\Delta_{\rm ent}^{-4/3}.
  \label{eq:gw_entropy_dilution_appendix}
\end{equation}
This factor is not applied separately to the baryon asymmetry.  The variables
in Sec.~\ref{sec:pbh_dirac} are normalized to the fixed initial density
$n_{\rm ref}$, and Eq.~\eqref{eq:etaB_observable} already converts the evolved
charge to the final baryon-to-photon ratio using $a_{\rm end}$ and
$g_{\star s}(T_{\rm end})$.  An additional division by
$\Delta_{\rm ent}$ would therefore count the same entropy dilution twice.

For a PBH-dominated epoch, a second characteristic scale is the horizon at
evaporation,
\begin{equation}
  f_{\rm ev}\simeq1.65\times10^{-5}\,{\rm Hz}
  \left(\frac{T_{\rm ev}}{100\,{\rm GeV}}\right)
  \left(\frac{g_\star(T_{\rm ev})}{106.75}\right)^{1/6},
  \qquad T_{\rm ev}\propto M_{\rm PBH}^{-3/2}.
  \label{eq:gw_evaporation_frequency}
\end{equation}
A formation-scale peak near Eq.~\eqref{eq:gw_frequency_mass_relation} and an
evaporation-scale feature near Eq.~\eqref{eq:gw_evaporation_frequency} would
probe complementary stages of the PBH history.  Increasing $\beta'$ can
strengthen GW production associated with a PBH-dominated epoch, while the
entropy release suppresses backgrounds generated earlier; the resulting
amplitude need not vary monotonically with $\beta'$.

Topological defects provide a more model-dependent probe.  If an ultraviolet
completion of the charged-mediator sector contains local strings formed at a
scale $v_s$, then~\cite{Siemens:2006yp,Auclair:2019wcv}
\begin{equation}
  G\mu\simeq2\pi B(\kappa)\left(\frac{v_s}{M_{\rm Pl}}\right)^2,
  \qquad \Omega_{\rm GW}^{\rm str}(f)\propto(G\mu)^2\mathcal S(f),
  \label{eq:string_gw_scaling}
\end{equation}
where $\mu$ is the string tension, $B(\kappa)$ is an order-one function of the
scalar-to-gauge mass ratio, and $\mathcal S(f)$ contains the loop distribution
and expansion history.  Writing $M_X\sim y_Xv_s$, one obtains the indicative
range
\begin{equation}
  G\mu\sim4\times10^{-26}y_X^{-2}
  \left(\frac{M_X}{10^6\,{\rm GeV}}\right)^2
  \;{\rm to}\;
  4\times10^{-20}y_X^{-2}
  \left(\frac{M_X}{10^9\,{\rm GeV}}\right)^2.
  \label{eq:string_tension_benchmark}
\end{equation}
This estimate is conditional on the assumed symmetry-breaking realization and
is not a prediction of the minimal Boltzmann system.  Unstable domain walls
have similarly been discussed as probes of high-scale Dirac
leptogenesis~\cite{Barman:2022diracGW}.

A blue-tilted primordial tensor spectrum is another possible high-frequency
source.  Schematically~\cite{Fujita:2014hha,Fujita:2018ehq,Barman:2024slw},
\begin{equation}
  \Omega_{\rm GW,0}^{\rm prim}(f)
  \simeq\frac{\Omega_{r,0}}{24}rA_s
  \left(\frac{f}{f_{\rm CMB}}\right)^{n_t},
  \label{eq:blue_tensor_scaling}
\end{equation}
where $r$ is the tensor-to-scalar ratio, $A_s$ is the scalar amplitude, $n_t$
is the tensor tilt, and $f_{\rm CMB}$ is the CMB pivot frequency, up to
transfer-function and reheating corrections.  Such a signal would probe the
small-scale inflationary dynamics that may generate the PBHs, but would not by
itself establish PBH-assisted leptogenesis.

The gravitational source can accompany either Dirac or Majorana PBH-assisted
leptogenesis, but the baryon asymmetries cannot be compared through parent
particle yields alone.  A useful formal representation is
\begin{equation}
  \eta_B^{A}
  =\int d\ln a\,
  \mathcal K_A(a;M_{\rm PBH},\beta',\boldsymbol\theta_A)\,
  \mathcal S_A(a;M_{\rm PBH},\beta',\boldsymbol\theta_A),
  \qquad A\in\{{\rm Dirac},{\rm Maj}\},
  \label{eq:gw_dirac_majorana_comparison}
\end{equation}
where $\boldsymbol\theta_A$ denotes the particle-physics parameters of the
chosen model, $\mathcal S_A$ is its time-dependent CP source, and
$\mathcal K_A$ denotes the corresponding response kernel, including washout,
sphaleron conversion, and the final normalization to photons.  For the Dirac
case, $\mathcal S_A$ contains Eq.~\eqref{eq:gw_dirac_source} and the kernel is
generated by the coupled system in Eq.~\eqref{eq:pbh_odd_system}.  The
Majorana source and kernel instead follow from heavy-neutrino production,
lepton-number-violating decays, and the associated washout processes.  A
GW-inferred $(M_{\rm PBH},\beta')$ may therefore be used as a common
cosmological input, but each baryon asymmetry must be obtained from its own
Boltzmann evolution~\cite{JyotiDas:2021shi,Borah:2024pbhaxion,Borah:2024memburden,Borah:2026zbl}.
The additional $\Delta N_{\rm eff}$ contribution from the populated
right-handed-neutrino sector remains a characteristic correlated observable
of the Dirac realization considered here.

\section*{Data availability}
The numerical results were obtained from the coupled equations and benchmark
inputs specified in the manuscript.  The implementation builds on the public
ULYSSES framework~\cite{Granelli:2020pim,Granelli:2023vcm,Granelli:2026goh}.
The numerical data supporting the findings of this work are not publicly
available but can be obtained from the authors upon reasonable request.

\section*{Acknowledgements}
The authors thank Julian Heeck, Jan Heisig, Nayan Das and Yuber Ferney Perez
Gonzalez for helpful comments on the manuscript. A.~G. acknowledges support
from the Royal Society, UK, funding reference: NIF\ R1\ 253963.
M.M.~acknowledges financial support from the research
project TAsP (Theoretical Astroparticle Physics), funded by the Istituto
Nazionale di Fisica Nucleare (INFN).

\bibliographystyle{apsrev4-1}

\bibliography{ref}

@article{MacGibbon:1991tj,
    author = "MacGibbon, Jane H.",
    title = "{Quark and gluon jet emission from primordial black holes. 2. The Lifetime emission}",
    reportNumber = "LHEA-91-001",
    doi = "10.1103/PhysRevD.44.376",
    journal = "Phys. Rev. D",
    volume = "44",
    pages = "376--392",
    year = "1991"
}

@article{MacGibbon:1990zk,
    author = "MacGibbon, J. H. and Webber, B. R.",
    title = "{Quark and gluon jet emission from primordial black holes: The instantaneous spectra}",
    doi = "10.1103/PhysRevD.41.3052",
    journal = "Phys. Rev. D",
    volume = "41",
    pages = "3052--3079",
    year = "1990"
}

@article{Carr:1974nx,
    author = "Carr, Bernard J. and Hawking, S. W.",
    title = "{Black holes in the early Universe}",
    doi = "10.1093/mnras/168.2.399",
    journal = "Mon. Not. Roy. Astron. Soc.",
    volume = "168",
    pages = "399--415",
    year = "1974"
}

@article{Carr:1975qj,
    author = "Carr, Bernard J.",
    title = "{The Primordial black hole mass spectrum}",
    doi = "10.1086/153853",
    journal = "Astrophys. J.",
    volume = "201",
    pages = "1--19",
    year = "1975"
}

@article{Carr:1976zz,
    author = "Carr, Bernard J.",
    title = "{Some cosmological consequences of primordial black-hole evaporations}",
    doi = "10.1086/154351",
    journal = "Astrophys. J.",
    volume = "206",
    pages = "8--25",
    year = "1976"
}

@article{Press:1973iz,
    author = "Press, William H. and Schechter, Paul",
    title = "{Formation of galaxies and clusters of galaxies by selfsimilar gravitational condensation}",
    doi = "10.1086/152650",
    journal = "Astrophys. J.",
    volume = "187",
    pages = "425--438",
    year = "1974"
}

@article{Villanueva-Domingo:2021spv,
    author = "Villanueva-Domingo, Pablo and Mena, Olga and Palomares-Ruiz, Sergio",
    title = "{A brief review on primordial black holes as dark matter}",
    eprint = "2103.12087",
    archivePrefix = "arXiv",
    primaryClass = "astro-ph.CO",
    doi = "10.3389/fspas.2021.681084",
    journal = "Front. Astron. Space Sci.",
    volume = "8",
    pages = "87",
    year = "2021"
}

@article{Sasaki:2018dmp,
    author = "Sasaki, Misao and Suyama, Teruaki and Tanaka, Takahiro and Yokoyama, Shuichiro",
    title = "{Primordial black holes\textemdash{}perspectives in gravitational wave astronomy}",
    eprint = "1801.05235",
    archivePrefix = "arXiv",
    primaryClass = "astro-ph.CO",
    doi = "10.1088/1361-6382/aaa7b4",
    journal = "Class. Quant. Grav.",
    volume = "35",
    number = "6",
    pages = "063001",
    year = "2018"
}

@article{Carr:2009jm,
    author = "Carr, B. J. and Kohri, Kazunori and Sendouda, Yuuiti and Yokoyama, Jun'ichi",
    title = "{New cosmological constraints on primordial black holes}",
    eprint = "0912.5297",
    archivePrefix = "arXiv",
    primaryClass = "astro-ph.CO",
    reportNumber = "RESCEU-31-09, TU-852, YITP-09-112",
    doi = "10.1103/PhysRevD.81.104019",
    journal = "Phys. Rev. D",
    volume = "81",
    pages = "104019",
    year = "2010"
}

@article{Page:1976df,
    author = "Page, Don N.",
    title = "{Particle Emission Rates from a Black Hole: Massless Particles from an Uncharged, Nonrotating Hole}",
    doi = "10.1103/PhysRevD.13.198",
    journal = "Phys. Rev. D",
    volume = "13",
    pages = "198--206",
    year = "1976"
}

@article{Page:1977um,
    author = "Page, Don N.",
    title = "{Particle Emission Rates from a Black Hole. 3. Charged Leptons from a Nonrotating Hole}",
    doi = "10.1103/PhysRevD.16.2402",
    journal = "Phys. Rev. D",
    volume = "16",
    pages = "2402--2411",
    year = "1977"
}

@article{Fujita:2014hha,
    author = "Fujita, Tomohiro and Kawasaki, Masahiro and Harigaya, Keisuke and Matsuda, Ryo",
    title = "{Baryon asymmetry, dark matter, and density perturbation from primordial black holes}",
    eprint = "1401.1909",
    archivePrefix = "arXiv",
    primaryClass = "astro-ph.CO",
    reportNumber = "IPMU-14-0009, ICRR-REPORT-668-2013-17",
    doi = "10.1103/PhysRevD.89.103501",
    journal = "Phys. Rev. D",
    volume = "89",
    number = "10",
    pages = "103501",
    year = "2014"
}

@article{Gunn:2024xaq,
    author = "Gunn, Jacob and Heurtier, Lucien and Perez-Gonzalez, Yuber F. and Turner, Jessica",
    title = "{Primordial black hole hot spots and out-of-equilibrium dynamics}",
    eprint = "2409.02173",
    archivePrefix = "arXiv",
    primaryClass = "hep-ph",
    reportNumber = "KCL-PH-TH-2024-48, IPPP/24/58",
    doi = "10.1088/1475-7516/2025/02/040",
    journal = "JCAP",
    volume = "02",
    pages = "040",
    year = "2025"
}

@article{Cheek:2021odj,
    author = "Cheek, Andrew and Heurtier, Lucien and Perez-Gonzalez, Yuber F. and Turner, Jessica",
    title = "{Primordial Black Hole Evaporation and Dark Matter Production: I. Solely Hawking radiation}",
    eprint = "2107.00013",
    archivePrefix = "arXiv",
    primaryClass = "hep-ph",
    doi = "10.1103/PhysRevD.105.015022",
    journal = "Phys. Rev. D",
    volume = "105",
    number = "1",
    pages = "015022",
    year = "2022"
}

@article{Cheek:2022dbx,
    author = "Cheek, Andrew and Heurtier, Lucien and Perez-Gonzalez, Yuber F. and Turner, Jessica",
    title = "{Redshift effects in particle production from Kerr primordial black holes}",
    eprint = "2207.09462",
    archivePrefix = "arXiv",
    primaryClass = "astro-ph.CO",
    doi = "10.1103/PhysRevD.106.103012",
    journal = "Phys. Rev. D",
    volume = "106",
    number = "10",
    pages = "103012",
    year = "2022"
}

@article{Calabrese:2023key,
    author = "Calabrese, Roberta and Chianese, Marco and Gunn, Jacob and Miele, Gennaro and Morisi, Stefano and Saviano, Ninetta",
    title = "{Limits on light primordial black holes from high-scale leptogenesis}",
    eprint = "2305.13369",
    archivePrefix = "arXiv",
    primaryClass = "hep-ph",
    doi = "10.1103/PhysRevD.107.123537",
    journal = "Phys. Rev. D",
    volume = "107",
    number = "12",
    pages = "123537",
    year = "2023"
}

@article{Perez-Gonzalez:2020vnz,
    author = "Perez-Gonzalez, Yuber F. and Turner, Jessica",
    title = "{Assessing the tension between a black hole dominated early universe and leptogenesis}",
    eprint = "2010.03565",
    archivePrefix = "arXiv",
    primaryClass = "hep-ph",
    reportNumber = "FERMILAB-PUB-20-528-T, NUHEP-TH/20-10, IPPP/20/46",
    doi = "10.1103/PhysRevD.104.103021",
    journal = "Phys. Rev. D",
    volume = "104",
    number = "10",
    pages = "103021",
    year = "2021"
}

@article{Dick:1999je,
    author = "Dick, Karin and Lindner, Manfred and Ratz, Michael and Wright, David",
    title = "{Leptogenesis with Dirac neutrinos}",
    eprint = "hep-ph/9907562",
    archivePrefix = "arXiv",
    doi = "10.1103/PhysRevLett.84.4039",
    journal = "Phys. Rev. Lett.",
    volume = "84",
    pages = "4039--4042",
    year = "2000"
}

@article{Murayama:2002je,
    author = "Murayama, Hitoshi and Pierce, Aaron",
    title = "{Realistic Dirac leptogenesis}",
    eprint = "hep-ph/0206177",
    archivePrefix = "arXiv",
    reportNumber = "UCB-PTH-02-26, LBNL-50854",
    doi = "10.1103/PhysRevLett.89.271601",
    journal = "Phys. Rev. Lett.",
    volume = "89",
    pages = "271601",
    year = "2002"
}

@article{Heeck:2023soj,
    author = "Heeck, Julian and Heisig, Jan and Thapa, Anil",
    title = "{Testing Dirac leptogenesis with the cosmic microwave background and proton decay}",
    eprint = "2304.09893",
    archivePrefix = "arXiv",
    primaryClass = "hep-ph",
    doi = "10.1103/PhysRevD.108.035014",
    journal = "Phys. Rev. D",
    volume = "108",
    number = "3",
    pages = "035014",
    year = "2023"
}

@article{DOnofrio:2014rug,
    author = "D'Onofrio, Michela and Rummukainen, Kari and Tranberg, Anders",
    title = "{Sphaleron Rate in the Minimal Standard Model}",
    eprint = "1404.3565",
    archivePrefix = "arXiv",
    primaryClass = "hep-ph",
    doi = "10.1103/PhysRevLett.113.141602",
    journal = "Phys. Rev. Lett.",
    volume = "113",
    number = "14",
    pages = "141602",
    year = "2014"
}

@article{Harvey:1990qw,
    author = "Harvey, Jeffrey A. and Turner, Michael S.",
    title = "{Cosmological baryon and lepton number in the presence of electroweak fermion number violation}",
    reportNumber = "FERMILAB-PUB-90-049-A, EFI-90-33",
    doi = "10.1103/PhysRevD.42.3344",
    journal = "Phys. Rev. D",
    volume = "42",
    pages = "3344--3349",
    year = "1990"
}

@article{PhysRevD.36.581,
  title = {Sphalerons, small fluctuations, and baryon-number violation in electroweak theory},
  author = {Arnold, Peter and McLerran, Larry},
  journal = {Phys. Rev. D},
  volume = {36},
  issue = {2},
  pages = {581--595},
  numpages = {0},
  year = {1987},
  month = {Jul},
  publisher = {American Physical Society},
  doi = {10.1103/PhysRevD.36.581},
  url = {https://link.aps.org/doi/10.1103/PhysRevD.36.581}
}

@article{Kuzmin:1985mm,
    author = "Kuzmin, V. A. and Rubakov, V. A. and Shaposhnikov, M. E.",
    title = "{On the Anomalous Electroweak Baryon Number Nonconservation in the Early Universe}",
    reportNumber = "IC/85/8",
    doi = "10.1016/0370-2693(85)91028-7",
    journal = "Phys. Lett. B",
    volume = "155",
    pages = "36",
    year = "1985"
}

@article{PhysRevD.30.2212,
  title = {A saddle-point solution in the Weinberg-Salam theory},
  author = {Klinkhamer, F. R. and Manton, N. S.},
  journal = {Phys. Rev. D},
  volume = {30},
  issue = {10},
  pages = {2212--2220},
  numpages = {0},
  year = {1984},
  month = {Nov},
  publisher = {American Physical Society},
  doi = {10.1103/PhysRevD.30.2212},
  url = {https://link.aps.org/doi/10.1103/PhysRevD.30.2212}
}

@article{Sakharov:1967dj,
  author       = {Sakharov, A. D.},
  title        = {Violation of {CP} Invariance, {C} Asymmetry, and Baryon Asymmetry of the Universe},
  journal      = {JETP Lett.},
  volume       = {5},
  pages        = {24--27},
  year         = {1967}
}

@article{Fukugita:1986hr,
  author       = {Fukugita, M. and Yanagida, T.},
  title        = {Baryogenesis Without Grand Unification},
  journal      = {Phys. Lett. B},
  volume       = {174},
  pages        = {45--47},
  year         = {1986},
  doi          = {10.1016/0370-2693(86)91126-3}
}

@article{Carr:2020gox,
  author       = {Carr, B. and Kohri, K. and Sendouda, Y. and Yokoyama, J.},
  title        = {Primordial Black Holes},
  journal      = {Rept. Prog. Phys.},
  volume       = {84},
  number       = {11},
  pages        = {116902},
  year         = {2021},
  eprint       = {2002.12778},
  archivePrefix= {arXiv},
  primaryClass = {astro-ph.CO},
  doi          = {10.1088/1361-6633/ac1e31}
}

@article{Hawking:1974rv,
  author       = {Hawking, S. W.},
  title        = {Black Hole Explosions?},
  journal      = {Nature},
  volume       = {248},
  pages        = {30--31},
  year         = {1974},
  doi          = {10.1038/248030a0}
}

@article{Hawking:1975vcx,
  author       = {Hawking, S. W.},
  title        = {Particle Creation by Black Holes},
  journal      = {Commun. Math. Phys.},
  volume       = {43},
  pages        = {199--220},
  year         = {1975},
  doi          = {10.1007/BF02345020}
}

@article{Gu:2006dc,
    author = "Gu, Pei-Hong and He, Hong-Jian",
    title = "{Neutrino Mass and Baryon Asymmetry from Dirac Seesaw}",
    eprint = "hep-ph/0610275",
    archivePrefix = "arXiv",
    doi = "10.1088/1475-7516/2006/12/010",
    journal = "JCAP",
    volume = "12",
    pages = "010",
    year = "2006"
}

@article{Gu:2007mc,
    author = "Gu, Pei-Hong and He, Hong-Jian and Sarkar, Utpal",
    title = "{Realistic neutrinogenesis with radiative vertex correction}",
    eprint = "0709.1019",
    archivePrefix = "arXiv",
    primaryClass = "hep-ph",
    doi = "10.1016/j.physletb.2007.11.061",
    journal = "Phys. Lett. B",
    volume = "659",
    pages = "634--639",
    year = "2008"
}

@article{Borah:2016zbd,
    author = "Borah, Debasish and Dasgupta, Arnab",
    title = "{Common Origin of Neutrino Mass, Dark Matter and Dirac Leptogenesis}",
    eprint = "1608.03872",
    archivePrefix = "arXiv",
    primaryClass = "hep-ph",
    doi = "10.1088/1475-7516/2016/12/034",
    journal = "JCAP",
    volume = "12",
    pages = "034",
    year = "2016"
}

@article{Narendra:2017uxl,
    author = "Narendra, Nimmala and Sahoo, Nirakar and Sahu, Narendra",
    title = "{Dark matter assisted Dirac leptogenesis and neutrino mass}",
    eprint = "1712.02960",
    archivePrefix = "arXiv",
    primaryClass = "hep-ph",
    doi = "10.1016/j.nuclphysb.2018.09.007",
    journal = "Nucl. Phys. B",
    volume = "936",
    pages = "76--90",
    year = "2018"
}

@article{Heeck:2013vha,
    author = "Heeck, Julian",
    title = "{Leptogenesis with Lepton-Number-Violating Dirac Neutrinos}",
    eprint = "1307.2241",
    archivePrefix = "arXiv",
    primaryClass = "hep-ph",
    doi = "10.1103/PhysRevD.88.076004",
    journal = "Phys. Rev. D",
    volume = "88",
    pages = "076004",
    year = "2013"
}

@article{Gu:2019yvw,
    author = "Gu, Pei-Hong",
    title = "{Leptogenesis with testable Dirac neutrino mass generation}",
    eprint = "1907.09443",
    archivePrefix = "arXiv",
    primaryClass = "hep-ph",
    doi = "10.1016/j.physletb.2020.135411",
    journal = "Phys. Lett. B",
    volume = "805",
    pages = "135411",
    year = "2020"
}

@article{Mahanta:2021plx,
    author = "Mahanta, Devabrat and Borah, Debasish",
    title = "{Low scale Dirac leptogenesis and dark matter with observable $\Delta N_{\mathrm{eff}}$}",
    eprint = "2101.02092",
    archivePrefix = "arXiv",
    primaryClass = "hep-ph",
    doi = "10.1140/epjc/s10052-022-10443-5",
    journal = "Eur. Phys. J. C",
    volume = "82",
    number = "5",
    pages = "495",
    year = "2022"
}

@article{Bernal:2022pue,
    author = "Bernal, Nicol{\'a}s and Fong, Chee Sheng and Perez-Gonzalez, Yuber F. and Turner, Jessica",
    title = "{Rescuing High-Scale Leptogenesis using Primordial Black Holes}",
    eprint = "2203.08823",
    archivePrefix = "arXiv",
    primaryClass = "hep-ph",
    reportNumber = "PI/UAN-2022-711FT, IPPP/22/12",
    doi = "10.1103/PhysRevD.106.035019",
    journal = "Phys. Rev. D",
    volume = "106",
    number = "3",
    pages = "035019",
    year = "2022"
}

@article{Planck:2018vyg,
    author = "Aghanim, N. and others",
    collaboration = "Planck",
    title = "{Planck 2018 results. VI. Cosmological parameters}",
    eprint = "1807.06209",
    archivePrefix = "arXiv",
    primaryClass = "astro-ph.CO",
    doi = "10.1051/0004-6361/201833910",
    journal = "Astron. Astrophys.",
    volume = "641",
    pages = "A6",
    year = "2020"
}

@misc{Abazajian:2019eic,
    author = "Abazajian, Kevork N. and others",
    title = "{CMB-S4 Science Case, Reference Design, and Project Plan}",
    eprint = "1907.04473",
    archivePrefix = "arXiv",
    primaryClass = "astro-ph.IM",
    year = "2019"
}

@article{Granelli:2020pim,
    author = "Granelli, Alessandro and Moffat, Kristian and Perez-Gonzalez, Yuber F. and Schulz, Holger and Turner, Jessica",
    title = "{ULYSSES: Universal LeptogeneSiS Equation Solver}",
    eprint = "2007.09150",
    archivePrefix = "arXiv",
    primaryClass = "hep-ph",
    reportNumber = "FERMILAB-PUB-20-275-T, SISSA 17/2020/FISI, IPPP/20/30",
    doi = "10.1016/j.cpc.2020.107813",
    journal = "Comput. Phys. Commun.",
    volume = "262",
    pages = "107813",
    year = "2021"
}

@article{Granelli:2023vcm,
    author = "Granelli, Alessandro and Leslie, Christopher and Perez-Gonzalez, Yuber F. and Schulz, Holger and Shuve, Brian and Turner, Jessica and Walker, Rosie",
    title = "{ULYSSES, universal LeptogeneSiS equation solver: Version 2}",
    eprint = "2301.05722",
    archivePrefix = "arXiv",
    primaryClass = "hep-ph",
    reportNumber = "IPPP/23/02",
    doi = "10.1016/j.cpc.2023.108834",
    journal = "Comput. Phys. Commun.",
    volume = "291",
    pages = "108834",
    year = "2023"
}

@article{Granelli:2026goh,
    author = "Granelli, Alessandro and Klari{\'c}, Juraj and Pasari, Dhruv and Perez-Gonzalez, Yuber F. and Turner, Jessica",
    title = "{ULYSSES the Third: An Odyssey Towards a Unified Python Toolkit for Leptogenesis}",
    eprint = "2605.16540",
    archivePrefix = "arXiv",
    primaryClass = "hep-ph",
    reportNumber = "IPPP/26/27, IFT-UAM/CSIC-26-64",
    month = "5",
    year = "2026"
}

@article{Harada:2013epa,
    author = "Harada, Tomohiro and Yoo, Chul-Moon and Kohri, Kazunori",
    title = "{Threshold of primordial black hole formation}",
    eprint = "1309.4201",
    archivePrefix = "arXiv",
    primaryClass = "astro-ph.CO",
    reportNumber = "RUP-13-9, KEK-COSMO-129, KEK-TH-1668",
    doi = "10.1103/PhysRevD.88.084051",
    journal = "Phys. Rev. D",
    volume = "88",
    number = "8",
    pages = "084051",
    year = "2013",
    note = "[Erratum: Phys.Rev.D 89, 029903 (2014)]"
}

@article{Musco:2018rwt,
    author = "Musco, Ilia",
    title = "{Threshold for primordial black holes: Dependence on the shape of the cosmological perturbations}",
    eprint = "1809.02127",
    archivePrefix = "arXiv",
    primaryClass = "gr-qc",
    doi = "10.1103/PhysRevD.100.123524",
    journal = "Phys. Rev. D",
    volume = "100",
    number = "12",
    pages = "123524",
    year = "2019"
}

@article{Morrison:2018xla,
    author = "Morrison, Logan and Profumo, Stefano and Yu, Yan",
    title = "{Melanopogenesis: Dark Matter of (almost) any Mass and Baryonic Matter from the Evaporation of Primordial Black Holes weighing a Ton (or less)}",
    eprint = "1812.10606",
    archivePrefix = "arXiv",
    primaryClass = "astro-ph.CO",
    doi = "10.1088/1475-7516/2019/05/005",
    journal = "JCAP",
    volume = "05",
    pages = "005",
    year = "2019"
}

@article{JyotiDas:2021shi,
    author = "Jyoti Das, Suruj and Mahanta, Devabrat and Borah, Debasish",
    title = "{Low scale leptogenesis and dark matter in the presence of primordial black holes}",
    eprint = "2104.14496",
    archivePrefix = "arXiv",
    primaryClass = "hep-ph",
    doi = "10.1088/1475-7516/2021/11/019",
    journal = "JCAP",
    volume = "11",
    pages = "019",
    year = "2021"
}

@article{Barman:2022gjo,
    author = "Barman, Basabendu and Borah, Debasish and Das Jyoti, Suruj and Roshan, Rishav",
    title = "{Cogenesis of Baryon asymmetry and gravitational dark matter from primordial black holes}",
    eprint = "2204.10339",
    archivePrefix = "arXiv",
    primaryClass = "hep-ph",
    reportNumber = "PI/UAN-2022-714FT",
    doi = "10.1088/1475-7516/2022/08/068",
    journal = "JCAP",
    volume = "08",
    pages = "068",
    year = "2022"
}

@article{Schmitz:2023pfy,
    author = "Schmitz, Kai and Xu, Xun-Jie",
    title = "{Wash-in leptogenesis after the evaporation of primordial black holes}",
    eprint = "2311.01089",
    archivePrefix = "arXiv",
    primaryClass = "hep-ph",
    reportNumber = "MS-TP-23-46",
    doi = "10.1016/j.physletb.2024.138473",
    journal = "Phys. Lett. B",
    volume = "849",
    pages = "138473",
    year = "2024"
}

@article{Calabrese:2023bxz,
    author = "Calabrese, Roberta and Chianese, Marco and Gunn, Jacob and Miele, Gennaro and Morisi, Stefano and Saviano, Ninetta",
    title = "{Impact of primordial black holes on heavy neutral leptons searches in the framework of resonant leptogenesis}",
    eprint = "2311.13276",
    archivePrefix = "arXiv",
    primaryClass = "hep-ph",
    doi = "10.1103/PhysRevD.109.103001",
    journal = "Phys. Rev. D",
    volume = "109",
    number = "10",
    pages = "103001",
    year = "2024"
}

@article{Ghoshal:2023fno,
    author = "Ghoshal, Anish and Perez-Gonzalez, Yuber F. and Turner, Jessica",
    title = "{Superradiant leptogenesis}",
    eprint = "2312.06768",
    archivePrefix = "arXiv",
    primaryClass = "hep-ph",
    reportNumber = "IPPP/23/78",
    doi = "10.1007/JHEP02(2024)113",
    journal = "JHEP",
    volume = "02",
    pages = "113",
    year = "2024"
}

@article{Barman:2024slw,
    author = "Barman, Basabendu and Jyoti Das, Suruj and Haque, Md Riajul and Mambrini, Yann",
    title = "{Leptogenesis, primordial gravitational waves, and PBH-induced reheating}",
    eprint = "2403.05626",
    archivePrefix = "arXiv",
    primaryClass = "hep-ph",
    reportNumber = "CTPU-PTC-24-07",
    doi = "10.1103/PhysRevD.110.043528",
    journal = "Phys. Rev. D",
    volume = "110",
    number = "4",
    pages = "043528",
    year = "2024"
}

@article{Calabrese:2025sfh,
    author = "Calabrese, Roberta and Chianese, Marco and Saviano, Ninetta",
    title = "{Impact of memory-burdened primordial black holes on high-scale leptogenesis}",
    eprint = "2501.06298",
    archivePrefix = "arXiv",
    primaryClass = "hep-ph",
    doi = "10.1103/PhysRevD.111.083008",
    journal = "Phys. Rev. D",
    volume = "111",
    number = "8",
    pages = "083008",
    year = "2025"
}

@article{Borah:2026zbl,
    author = "Borah, Debasish and Das, Nayan",
    title = "{Multi-peaked high-frequency gravitational waves from PBH-assisted leptogenesis}",
    eprint = "2606.23787",
    archivePrefix = "arXiv",
    primaryClass = "hep-ph",
    month = "6",
    year = "2026"
}

@article{Turner:1979bt,
    author = "Turner, Michael S.",
    title = "{BARYON PRODUCTION BY PRIMORDIAL BLACK HOLES}",
    reportNumber = "EFI-79/45-CHICAGO",
    doi = "10.1016/0370-2693(79)90095-9",
    journal = "Phys. Lett. B",
    volume = "89",
    pages = "155--159",
    year = "1979"
}

@article{Baumann:2007yr,
    author = "Baumann, Daniel and Steinhardt, Paul J. and Turok, Neil",
    title = "{Primordial Black Hole Baryogenesis}",
    eprint = "hep-th/0703250",
    archivePrefix = "arXiv",
    reportNumber = "PUPT-2229",
    month = "3",
    year = "2007"
}

@article{Blazek:2024efd,
    author = "Bla{\v{z}}ek, Tom{\'a}{\v{s}} and Heeck, Julian and Heisig, Jan and Mat{\'a}k, Peter and Zaujec, Viktor",
    title = "{Dirac leptogenesis from asymmetry wash-in via scatterings}",
    eprint = "2404.16934",
    archivePrefix = "arXiv",
    primaryClass = "hep-ph",
    reportNumber = "TTK-24-14",
    doi = "10.1103/PhysRevD.110.055042",
    journal = "Phys. Rev. D",
    volume = "110",
    number = "5",
    pages = "055042",
    year = "2024"
}

@article{Hooper:2020otu,
    author = "Hooper, Dan and Krnjaic, Gordan",
    title = "{GUT Baryogenesis With Primordial Black Holes}",
    eprint = "2010.01134",
    archivePrefix = "arXiv",
    primaryClass = "hep-ph",
    reportNumber = "FERMILAB-PUB-21-144-AE",
    doi = "10.1103/PhysRevD.103.043504",
    journal = "Phys. Rev. D",
    volume = "103",
    number = "4",
    pages = "043504",
    year = "2021"
}

@article{deSalas:2016ztq,
    author = "de Salas, Pablo F. and Pastor, Sergio",
    title = "{Relic neutrino decoupling with flavour oscillations revisited}",
    eprint = "1606.06986",
    archivePrefix = "arXiv",
    primaryClass = "hep-ph",
    doi = "10.1088/1475-7516/2016/07/051",
    journal = "JCAP",
    volume = "07",
    pages = "051",
    year = "2016"
}

@article{Ananda:2006af,
  author = {Ananda, Kishore N. and Clarkson, Chris and Wands, David},
  title = {The cosmological gravitational wave background from primordial density perturbations},
  eprint = {gr-qc/0612013},
  archivePrefix = {arXiv},
  doi = {10.1103/PhysRevD.75.123518},
  journal = {Phys. Rev. D},
  volume = {75},
  pages = {123518},
  year = {2007}
}

@article{Baumann:2007zm,
  author = {Baumann, Daniel and Steinhardt, Paul and Takahashi, Keitaro and Ichiki, Kiyotomo},
  title = {Gravitational Wave Spectrum Induced by Primordial Scalar Perturbations},
  eprint = {hep-th/0703290},
  archivePrefix = {arXiv},
  doi = {10.1103/PhysRevD.76.084019},
  journal = {Phys. Rev. D},
  volume = {76},
  pages = {084019},
  year = {2007}
}

@article{Inomata:2016rbd,
  author = {Inomata, Keisuke and Kawasaki, Masahiro and Mukaida, Kyohei and Tada, Yuichiro and Yanagida, Tsutomu T.},
  title = {Inflationary primordial black holes for the LIGO gravitational wave events and pulsar timing array experiments},
  eprint = {1611.06130},
  archivePrefix = {arXiv},
  primaryClass = {astro-ph.CO},
  doi = {10.1103/PhysRevD.95.123510},
  journal = {Phys. Rev. D},
  volume = {95},
  number = {12},
  pages = {123510},
  year = {2017}
}

@article{Kohri:2018awv,
  author = {Kohri, Kazunori and Terada, Takahiro},
  title = {Semianalytic calculation of gravitational wave spectrum nonlinearly induced from primordial curvature perturbations},
  eprint = {1804.08577},
  archivePrefix = {arXiv},
  primaryClass = {gr-qc},
  doi = {10.1103/PhysRevD.97.123532},
  journal = {Phys. Rev. D},
  volume = {97},
  number = {12},
  pages = {123532},
  year = {2018}
}

@article{Inomata:2020lmk,
  author = {Inomata, Keisuke and Kawasaki, Masahiro and Mukaida, Kyohei and Terada, Takahiro and Yanagida, Tsutomu T.},
  title = {Gravitational Wave Production right after a Primordial Black Hole Evaporation},
  eprint = {2003.10455},
  archivePrefix = {arXiv},
  primaryClass = {astro-ph.CO},
  doi = {10.1103/PhysRevD.101.123533},
  journal = {Phys. Rev. D},
  volume = {101},
  number = {12},
  pages = {123533},
  year = {2020}
}

@article{Siemens:2006yp,
  author = {Siemens, Xavier and Mandic, Vuk and Creighton, Jolien},
  title = {Gravitational wave stochastic background from cosmic strings},
  eprint = {astro-ph/0610920},
  archivePrefix = {arXiv},
  doi = {10.1103/PhysRevLett.98.111101},
  journal = {Phys. Rev. Lett.},
  volume = {98},
  pages = {111101},
  year = {2007}
}

@article{Auclair:2019wcv,
  author = {Auclair, Pierre and others},
  title = {Probing the gravitational wave background from cosmic strings with LISA},
  eprint = {1909.00819},
  archivePrefix = {arXiv},
  primaryClass = {astro-ph.CO},
  doi = {10.1088/1475-7516/2020/04/034},
  journal = {JCAP},
  volume = {04},
  pages = {034},
  year = {2020}
}

@article{Fujita:2018ehq,
  author = {Fujita, Tomohiro and Kuroyanagi, Sachiko and Mizuno, Shuntaro and Mukohyama, Shinji},
  title = {Blue-tilted Primordial Gravitational Waves from Massive Gravity},
  eprint = {1808.02381},
  archivePrefix = {arXiv},
  primaryClass = {gr-qc},
  doi = {10.1016/j.physletb.2018.12.025},
  journal = {Phys. Lett. B},
  volume = {789},
  pages = {215--219},
  year = {2019}
}

@article{Barman:2022diracGW,
  author = {Barman, Basabendu and Borah, Debasish and Dasgupta, Arnab and Ghoshal, Anish},
  title = {Probing High Scale Dirac Leptogenesis via Gravitational Waves from Domain Walls},
  eprint = {2205.03422},
  archivePrefix = {arXiv},
  primaryClass = {hep-ph},
  doi = {10.1103/PhysRevD.106.015007},
  journal = {Phys. Rev. D},
  volume = {106},
  number = {1},
  pages = {015007},
  year = {2022}
}

@article{Borah:2024pbhaxion,
  author = {Borah, Debasish and Das, Nayan and Das, Suruj Jyoti and Samanta, Rome},
  title = {Cogenesis of baryon and dark matter with PBHs and the QCD axion},
  eprint = {2403.02401},
  archivePrefix = {arXiv},
  primaryClass = {hep-ph},
  doi = {10.1103/PhysRevD.110.115013},
  journal = {Phys. Rev. D},
  volume = {110},
  pages = {115013},
  year = {2024}
}

@article{Borah:2024memburden,
  author = {Borah, Debasish and Das, Nayan},
  title = {Successful cogenesis of baryon and dark matter from memory-burdened PBH},
  eprint = {2410.16403},
  archivePrefix = {arXiv},
  primaryClass = {hep-ph},
  doi = {10.1088/1475-7516/2025/02/031},
  journal = {JCAP},
  volume = {02},
  pages = {031},
  year = {2025}
}

@article{Babu:2024lrsm,
 author = {Babu, K. S. and Kaladharan, Ajay},
 title = {Dirac Leptogenesis in Left-Right Symmetric Models},
 eprint = {2410.24125},
 archivePrefix = {arXiv},
 primaryClass = {hep-ph},
 doi = {10.1103/kxvb-l4gb},
 journal = {Phys. Rev. D},
 volume = {112},
 pages = {035015},
 year = {2025}
}

@article{Ahmed:2025primordial,
 author = {Ahmed, Aqeel and Garc\'es, Juan P. and Lindner, Manfred},
 title = {Primordial Dirac Leptogenesis},
 eprint = {2511.03794},
 archivePrefix = {arXiv},
 primaryClass = {hep-ph},
 doi = {10.1103/bnfj-ypx8},
 journal = {Phys. Rev. D},
 volume = {113},
 pages = {L111702},
 year = {2026}
}

@article{Cheek:2022massspin,
 author = {Cheek, Andrew and Heurtier, Lucien and Perez-Gonzalez, Yuber F. and Turner, Jessica},
 title = {Evaporation of Primordial Black Holes in the Early Universe: Mass and Spin Distributions},
 eprint = {2212.03878},
 archivePrefix = {arXiv},
 primaryClass = {hep-ph},
 doi = {10.1103/PhysRevD.108.015005},
 journal = {Phys. Rev. D},
 volume = {108},
 pages = {015005},
 year = {2023}
}

@article{Ishida:2025adm,
 author = {Ishida, Megumi and Ohki, Hiroshi and Uemura, Shohei},
 title = {Unified Origin of Dirac Neutrino and Asymmetric Dark Matter Masses via a Dirac-Type Leptogenesis},
 eprint = {2510.13723},
 archivePrefix = {arXiv},
 primaryClass = {hep-ph},
 doi = {10.1093/ptep/ptaf185},
 journal = {Prog. Theor. Exp. Phys.},
 volume = {2026},
 number = {1},
 pages = {013B08},
 year = {2026}
}

@article{Barman:2023colored,
 author = {Barman, Basabendu and Borah, Debasish and Das, Suruj Jyoti and Roshan, Rishav},
 title = {Gravitational wave signatures of a PBH-generated baryon-dark matter coincidence},
 eprint = {2212.00052},
 archivePrefix = {arXiv},
 primaryClass = {hep-ph},
 doi = {10.1103/PhysRevD.107.095002},
 journal = {Phys. Rev. D},
 volume = {107},
 pages = {095002},
 year = {2023}
}

@article{Iguaz:2025asymmetric,
 author = {Iguaz Juan, Joaquim and Perez-Gonzalez, Yuber F. and Turner, Jessica},
 title = {Baryogenesis via Asymmetric Evaporation of Primordial Black Holes},
 eprint = {2508.21011},
 archivePrefix = {arXiv},
 primaryClass = {hep-ph},
 doi = {10.1088/1475-7516/2026/05/055},
 journal = {JCAP},
 volume = {05},
 pages = {055},
 year = {2026}
}

@article{Goldstein:2026iuu,
    author = "Goldstein, Samuel and Hill, J. Colin",
    title = "{2{\\%} determination of Neff from primordial element abundance, cosmic microwave background, and baryon acoustic oscillation measurements}",
    eprint = "2603.13226",
    archivePrefix = "arXiv",
    primaryClass = "astro-ph.CO",
    doi = "10.1103/5vk8-v1rb",
    journal = "Phys. Rev. D",
    volume = "114",
    number = "2",
    pages = "L021305",
    year = "2026"
}

\end{document}